\documentclass[twocolumn,twocolappendix]{aastex63}
\usepackage[T1]{fontenc}
\usepackage{ae,aecompl}

\usepackage{graphicx}   
\usepackage{amsmath}    
\usepackage{amssymb}    

\newcommand{\figureref}[1]{Fig.~\ref{#1}}

\newcommand{\sectionref}[1]{Section \ref{#1}}
\newcommand{\appendixref}[1]{Appendix \ref{#1}}

\received{XXX}
\revised{XXX}
\accepted{XXX}
\submitjournal{AJ}
\shorttitle{Polarized thermal dust emission}
\shortauthors{Vandenbroucke et al.}

\begin{document}

\title{CosTuuM: polarized thermal dust emission by magnetically oriented
spheroidal grains}

\correspondingauthor{Bert Vandenbroucke}
\email{bert.vandenbroucke@ugent.be}

\author[0000-0001-7241-1704]{Bert Vandenbroucke}
\affiliation{Sterrenkundig Observatorium, Universiteit Gent, Krijgslaan 281,
B-9000 Gent, Belgium}
\author[0000-0002-3930-2757]{Maarten Baes}
\affiliation{Sterrenkundig Observatorium, Universiteit Gent, Krijgslaan 281,
B-9000 Gent, Belgium}
\author[0000-0002-4479-4119]{Peter Camps}
\affiliation{Sterrenkundig Observatorium, Universiteit Gent, Krijgslaan 281,
B-9000 Gent, Belgium}

\begin{abstract}

We present the new open source C++-based Python library \textsc{CosTuuM} 
that can be used to generate infrared absorption and emission 
coefficients for arbitrary mixtures of spheroidal dust grains that are 
(partially) aligned with a magnetic field. We outline the algorithms 
underlying the software, demonstrate the accuracy of our results using 
benchmarks from literature, and use our tool to investigate some 
commonly used approximative recipes. We find that the linear 
polarization fraction for a partially aligned dust grain mixture can be 
accurately represented by an appropriate linear combination of perfectly 
aligned grains and grains that are randomly oriented, but that the 
commonly used picket fence alignment breaks down for short wavelengths. 
We also find that for a fixed dust grain size, the 
absorption coefficients and linear polarization fraction for a 
realistic mixture of grains with various shapes cannot both 
be accurately represented by a single representative grain with a fixed 
shape, but that instead an average over an appropriate shape 
distribution should be used. Insufficient knowledge of an appropriate 
shape distribution is the main obstacle in obtaining accurate optical 
properties. \textsc{CosTuuM} is available as a standalone Python library 
and can be used to generate optical properties to be used in radiative 
transfer applications.

\end{abstract}


\section{Introduction}

The polarization of detected radiation from astrophysical sources offers an
additional window on some of the physical processes that happen in these
sources \citep{2009Matthews, 2019Andre}. A consistent polarization signature in
the radiation scattered off spherical dust grains e.g. contains information
about the spatial origin of that radiation. In emission, polarization traces the
orientation of non-spherical dust grains, as thermal emission from these grains
is preferentially polarized along the longest axis of the dust grain.

Dust grains can under some conditions align parallel or perpendicular to present
magnetic fields \citep[see][for a review]{2015Andersson}. This makes it possible
to observe the orientation of interstellar magnetic fields through polarimetry,
as was e.g. done by Planck \citep{2018Planck} for the Milky Way. More recently,
ALMA \citep{2016Cortes}, the POL-2/SCUBA2 polarimeter on the JCMT
\citep{2017Pattle}, and the HAWC+ polarimeter on the SOFIA telescope
\citep{2019Santos, 2020LopezRodriguez} have started opening up a new level of
magnetic field surveys by producing polarization maps for star forming molecular
clouds and external galaxies. Other polarimeters are being commissioned, e.g.
the TolTEC instrument on the Large Millimeter Telescope \citep{2018Bryan}, or
are being proposed, e.g. the B-BOP polarimeter aboard the ESA/JAXA mission SPICA
\citep{2018Roelfsema, 2019Andre}.

Computing the polarized emission properties of dust grains is a complex and
computationally challenging problem. Therefore, models of polarized dust
emission either focus on a limited number of alignment mechanisms, and/or make
strong assumptions about the degree of alignment (e.g. perfect alignment) and
the type and shape of aligned grains \citep[e.g.][]{2007Pelkonen, 2009Pelkonen,
2014Siebenmorgen, 2016Reissl, 2017Bertrang, 2019Hensley}. However, new planned
observational facilities offer a prospect of a much larger volume of
polarimetric data on various scales, and this prompts the development of more
advanced tools that can provide the material properties that are required to
accurately model polarization within forward radiation transfer (RT) modeling.
These tools will enable a rigorous comparison between the growing number of
models that include magnetohydrodynamics (MHD) and the increasing volume of
polarimetric data for the ISM over a large wavelength range, both on
protoplanetary disc scales \citep{2017BertrangFlockWolf, 2019Wurster}, the
scales of star-forming clouds and filaments \citep{2018Hennebelle,
2019Seifried}, as on galactic scales \citep{2017Grand, 2019Nelson,
2019Pillepich}.

Various methods and tools have been developed to compute emission properties
for spheroidal dust grains, e.g. the discrete dipole approximation (DDA) method
\citep{1994Draine}, the separation of variables method (SVM)
\citep{1993Voshchinnikov}, and the T-matrix method \citep{1971Waterman,
1998Mishchenko}. All these methods yield accurate results for small grains
and long wavelengths. For shorter wavelengths, the SVM and T-matrix methods
become very computationally demanding and run into numerical issues. The more
approximate DDA method can in principle handle arbitrary grain shapes and sizes,
yet at a relatively high computational cost \citep{1996Hovenier}.

In this paper, we introduce the open source software package
\textsc{CosTuuM}\footnote{CosTuuM is hosted in a public repository on
\url{https://github.com/SKIRT/CosTuuM}. Version 1.0 was archived on Zenodo under
doi:10.5281/zenodo.3842422.} (C++ T-Matrix method). \textsc{CosTuuM} was
primarily developed to provide material properties for the RT code SKIRT
\citep{2011Baes, 2015Camps, 2020Camps}, but is provided as a standalone package
that can generate material properties for an arbitrary mix of spheroidal dust
grains with various degrees of alignment. The tool is based on the T-matrix
method, and was benchmarked against the original Fortran code provided by
\citet{1998Mishchenko}. Unlike previous Python wrappers around the
\citet{1998Mishchenko} code \citep{2014Leinonen}, we have reimplemented the
T-matrix algorithm in modern object-oriented C++ using a state-of-the-art
task-based parallelisation strategy. This has allowed us to incorporate
additional physics, like grain alignment and grain shape distributions directly
into the library at increased efficiency. The library has been exposed to Python
and interoperates with the NumPy library.

The structure of the paper is as follows. In \sectionref{sec:method}, we
summarize the most important aspects of the T-matrix method and highlight the
conventions used in \textsc{CosTuuM}. We also introduce the necessary components
to appropriately average properties for a statistical mixture of dust grains,
like the alignment and shape distributions. In \sectionref{sec:results}, we
compare \textsc{CosTuuM} to existing benchmarks found in literature. We also use
\textsc{CosTuuM} to compute absorption coefficients and linear polarization
fractions for a range of dust grain models, and compare this with approximations
that are commonly made in literature. In \sectionref{sec:design}, we discuss our
software design and parallelisation strategy and give some practical examples of
using the tool. We conclude in \sectionref{sec:conclusion} with some general
conclusions. Details of the T-matrix method are presented in appendices.

\section{Method}
\label{sec:method}

For this work, we will assume that interstellar dust can be represented by a
mixture of spheroidal grains. The surface of these grains is described by the
equation
\begin{equation}
\frac{x^2}{b^2} + \frac{y^2}{b^2} + \frac{z^2}{c^2} = 1.
\end{equation}
The axis ratio, $d=b/c$, determines the shape of the spheroid; spheroids with
$d<1$ have two semi-minor axes and one semi-major axis and are \emph{prolate},
while spheroids with $d>1$ have two semi-major axes and one semi-minor axis and
are called \emph{oblate}. Spheroids with $d=1$ have three equal axes and hence
correspond to spheres.

It is common practice to refer to the size of a spheroidal dust grain in terms
of the size of a sphere with the same volume:
\begin{equation}
a = b^{2/3} c^{1/3} = b d^{-1/3}.
\end{equation}
Below we will parameterize spheroidal grains in terms of $a$ and $d$.

The material properties of a spheroidal dust grain with a specific composition
are encoded in its complex refractive index, $m_r = m_{r,r} + i m_{r,i}$, that
can be derived from the dielectric function for a material with that
composition.

For a dust grain with a given size ($a$), shape ($d$) and composition ($m_r$),
we want to compute the emission as a function of the outgoing angles $(\theta{},
\phi{})$ w.r.t. the symmetry axis $z$ of the spheroid. Because of azimuthal
symmetry, this emission will only depend on the zenith angle, $\theta{}$. When
discussing grains that are (partially) aligned with a magnetic field, we will
always assume that the magnetic field direction is along $z$ as well; in cases
where the symmetry axis of the grain is not perfectly aligned with the magnetic
field, appropriate transformations will take care of the change of reference
frame for the dust grain before alignment is discussed.

\subsection{Absorption coefficients}

\begin{figure}
\centering{}
\includegraphics[width=0.48\textwidth{}]{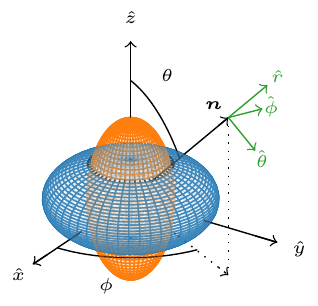}
\caption{Reference frame for our spheroidal grain model. An abritrary direction
$\boldsymbol{n}$ is characterized by a zenith angle $\theta{}$ and an azimuth
angle $\phi{}$, defined as indicated. For every direction, a spherical reference
frame $(\hat{r}, \hat{\theta{}}, \hat{\phi{}})$ can be defined as shown in
green. The symmetry axis of the spheroids is assumed to be along the $z$ axis,
for prolate spheroids (orange) this is the single long axis, for oblate
spheroids (blue) this is the single short axis.}
\label{fig:reference_frame}
\end{figure}

We represent the polarized radiation using the four element Stokes vector
(\citealp{2000Mishchenko_book}, but noting the convention difference explained
in \citealp{2017Peest}):
\begin{equation}
\mathbf{I} = \begin{pmatrix} I \\ Q \\ U \\ V \end{pmatrix} =
\begin{pmatrix}
E_{\theta{}} E^*_{\theta{}} + E_{\phi{}} E^*_{\phi{}} \\
E_{\theta{}} E^*_{\theta{}} - E_{\phi{}} E^*_{\phi{}} \\
E_{\theta{}} E^*_{\phi{}} + E_{\phi{}} E^*_{\theta{}} \\
i\left(E_{\phi{}} E^*_{\theta{}} - E_{\theta{}} E^*_{\phi{}}\right)
\end{pmatrix},
\end{equation}
where $^*$ represent the complex conjugate operation and we have assumed that
the (generally complex) transverse components of the electromagnetic wave are
expressed in a spherical right-handed coordinate frame $(\hat{r},
\hat{\theta{}}, \hat{\phi{}})$ with the radial direction along the propagation
direction of the wave, pointing away from the origin (see
\figureref{fig:reference_frame}). The vertical reference direction used to
compute $\theta{}$ is given by an arbitrary \emph{north} direction, which we
will always choose to be along our $z$ axis. As a direct corollary, we will only
have one linear polarization component, $Q$, while $U=0$ in all cases.

When an incoming plane electromagnetic wave with propagation direction
$\boldsymbol{n}_{\rm{}i}=(\theta{}_{\rm{}i}, \phi{}_{\rm{}i})$ scatters due to
an interaction with a spheroidal dust particle with an orientation
$\boldsymbol{n}_{\rm{}p}$ into a spherical wave with propagation direction
$\boldsymbol{n}_{\rm{}s}$, its Stokes vector $\mathbf{I}^{\rm{}i}$ is
converted into a new Stokes vector $\mathbf{I}^{\rm{}s}$. This new Stokes
vector can be related to the original Stokes vector by means of the M\"{u}ller
matrix $\mathbf{M}(\boldsymbol{n}_{\rm{}s},
\boldsymbol{n}_{\rm{}i}; \boldsymbol{n}_{\rm{}p})$:
\begin{equation}
\mathbf{I}^{\rm{}s} = \frac{1}{R^2} \mathbf{M}(\boldsymbol{n}_{\rm{}s},
\boldsymbol{n}_{\rm{}i}; \boldsymbol{n}_{\rm{}p}) \mathbf{I}^{\rm{}i},
\end{equation}
where $R$ represents the linear distance traveled by the wave after scattering.
Note that $\mathbf{M}$ has the dimensions of a surface area; it represents a
\emph{scattering cross section}.

Due to scattering, but also due to absorption of part of the electromagnetic
energy within the scattering particle, a beam of light moving over a distance
${\rm{}d}s$ in a direction $\boldsymbol{n}$ through a medium consisting of
spheroidal dust grains with the same orientation $\boldsymbol{n}_{\rm{}p}$
and density $n_{\rm{}p}$ will experience extinction, characterized by the
extinction matrix $\mathbf{K}(\boldsymbol{n}; \boldsymbol{n}_{\rm{}p})$:
\begin{equation}
\frac{{\rm{}d}\mathbf{I}}{{\rm{}d}s} = -n_{\rm{}p}
\mathbf{K}(\boldsymbol{n}; \boldsymbol{n}_{\rm{}p}) \mathbf{I}.
\end{equation}
Like the M\"{u}ller matrix, the extinction matrix has the dimensions of a
surface area.

To compute the absorption in a direction $\boldsymbol{n}$, we need to determine
the absorption cross sections $\mathbf{K}_{\rm{}a}(\boldsymbol{n};
\boldsymbol{n}_{\rm{}p})$ for the incoming Stokes vector components. This
requires subtracting the total contribution of incoming waves scattering into
the direction of interest from the relevant components of the extinction matrix:
\begin{equation}
K_{{\rm{}a},X}(\boldsymbol{n}; \boldsymbol{n}_{\rm{}p}) =
K_{XI}(\boldsymbol{n}; \boldsymbol{n}_{\rm{}p}) - \int M_{XI}(\boldsymbol{n},
\boldsymbol{n'}; \boldsymbol{n}_{\rm{}p}) {\rm{}d}\boldsymbol{n}',
\label{eq:absorption}
\end{equation}
where $X=I,Q,U,V$, and $M_{XI}$ represents the element of the matrix
$\mathbf{M}$ that relates the incoming Stokes vector component $X$ to the
outgoing Stokes vector component $I$.

In thermal equilibrium, the energy that is absorbed by the dust grains will be
emitted again at different wavelengths. In this case, the absorption for the
different Stokes vector components matches the corresponding emission, so once
the absorption cross sections are known, the emission cross sections are also
known.

The extinction and absorption cross sections are usually expressed as
dimensionless quantities (extinction/absorption coefficients) by dividing by the
cross section of a spherical grain with the same radius \citep{1991Mishchenko}:
\begin{align}
Q_{\rm{}ext} &= \frac{K_{II}}{\pi{}a^2},
&Q_{\rm{}ext,pol} = \frac{K_{QI}}{\pi{}a^2}, \label{eq:Qext} \\
Q_{\rm{}abs} &= \frac{K_{{\rm{}a},I}}{\pi{}a^2},
&Q_{\rm{}abs,pol} = \frac{K_{{\rm{}a},Q}}{\pi{}a^2}. \label{eq:Qabs}
\end{align}

Calculation of the absorption and emission coefficients for a specific
propagation direction hence requires the calculation of the extinction matrix
for that direction, as well as the calculation of the M\"{u}ller matrix for that
outgoing direction integrated over all incoming directions. Additionally, the
real interstellar medium does not consist of a single dust species with a fixed
orientation, so that in general we need to replace the extinction and M\"{u}ller
matrices in these expressions with suitably ensemble averaged versions for a
dust mixture with a specific size and shape distribution, and some prescribed
alignment distribution.

To compute the elements of the extinction and M\"{u}ller matrices, we need the
complex \emph{forward scattering matrix} $\mathbf{S}(\boldsymbol{n}_{\rm{}s},
\boldsymbol{n}_{\rm{}i}; \boldsymbol{n}_{\rm{}p})$ that links the transverse
components of the electromagnetic field before and after the scattering event:
\begin{equation}
\begin{pmatrix} E^{\rm{}s}_\theta{} \\ E^{\rm{}s}_\phi{} \end{pmatrix} =
\frac{1}{R} {\rm{}e}^{ikR} \mathbf{S}(\boldsymbol{n}_{\rm{}s},
\boldsymbol{n}_{\rm{}i}; \boldsymbol{n}_{\rm{}p}) \begin{pmatrix}
E^{\rm{}i}_\theta{} \\ E^{\rm{}i}_\phi{} \end{pmatrix},
\end{equation}
where $k=|\boldsymbol{k}|$ is the norm of the wavenumber vector. This matrix can
be calculated in various ways; we will use the so called T-matrix formalism (see
\appendixref{appendix:Tmatrix}), which is based on an expansion of the incoming
and outgoing electromagnetic waves in spherical basis functions. A powerful
aspect of this formalism is that it allows for the construction of a single
matrix $\mathbf{T}$ that encodes all properties for a dust mixture with a given
alignment distribution around some reference direction, from which a single
$\mathbf{S}$ matrix can be computed without the need for any additional
averaging over particle orientations.

Once the $\mathbf{S}$ matrix is known within some reference frame, the
extinction and M\"{u}ller matrices within that same reference frame can be
computed directly \citep{2000Mishchenko_book}:
\begin{align}
M_{II} &= \frac{1}{2} \left( |S_{\theta{}\theta{}}|^2 + |S_{\theta{}\phi{}}|^2 +
|S_{\phi{}\theta{}}|^2 + |S_{\phi{}\phi{}}|^2 \right), \label{eq:sca_start}\\
M_{IQ} &= \frac{1}{2} \left( |S_{\theta{}\theta{}}|^2 - |S_{\theta{}\phi{}}|^2 +
|S_{\phi{}\theta{}}|^2 - |S_{\phi{}\phi{}}|^2 \right),\\
M_{IU} &= {\rm{}Re}\left( S_{\theta{}\theta{}} S^*_{\theta{}\phi{}} +
S_{\phi{}\phi{}} S^*_{\phi{}\theta{}} \right),\\
M_{IV} &= -{\rm{}Im}\left( S_{\theta{}\theta{}} S^*_{\theta{}\phi{}} -
S_{\phi{}\phi{}} S^*_{\phi{}\theta{}} \right),
\end{align}
\begin{align}
M_{QI} &= \frac{1}{2} \left( |S_{\theta{}\theta{}}|^2 + |S_{\theta{}\phi{}}|^2 -
|S_{\phi{}\theta{}}|^2 - |S_{\phi{}\phi{}}|^2 \right),\\
M_{QQ} &= \frac{1}{2} \left( |S_{\theta{}\theta{}}|^2 - |S_{\theta{}\phi{}}|^2 -
|S_{\phi{}\theta{}}|^2 + |S_{\phi{}\phi{}}|^2 \right),\\
M_{QU} &= {\rm{}Re}\left( S_{\theta{}\theta{}} S^*_{\theta{}\phi{}} -
S_{\phi{}\phi{}} S^*_{\phi{}\theta{}} \right),\\
M_{QV} &= -{\rm{}Im}\left( S_{\theta{}\theta{}} S^*_{\theta{}\phi{}} +
S_{\phi{}\phi{}} S^*_{\phi{}\theta{}} \right),
\end{align}
\begin{align}
M_{UI} &= {\rm{}Re}\left( S_{\theta{}\theta{}} S^*_{\phi{}\theta{}} +
S_{\phi{}\phi{}} S^*_{\theta{}\phi{}} \right),\\
M_{UQ} &= {\rm{}Re}\left( S_{\theta{}\theta{}} S^*_{\phi{}\theta{}} -
S_{\phi{}\phi{}} S^*_{\theta{}\phi{}} \right),\\
M_{UU} &= {\rm{}Re}\left( S_{\theta{}\theta{}} S^*_{\phi{}\phi{}} +
S_{\theta{}\phi{}} S^*_{\phi{}\theta{}} \right),\\
M_{UV} &= -{\rm{}Im}\left( S_{\theta{}\theta{}} S^*_{\phi{}\phi{}} +
S_{\phi{}\theta{}} S^*_{\theta{}\phi{}} \right),
\end{align}
\begin{align}
M_{VI} &= -{\rm{}Im}\left( S_{\phi{}\theta{}} S^*_{\theta{}\theta{}} +
S_{\phi{}\phi{}} S^*_{\theta{}\phi{}} \right),\\
M_{VQ} &= -{\rm{}Im}\left( S_{\phi{}\theta{}} S^*_{\theta{}\theta{}} -
S_{\phi{}\phi{}} S^*_{\theta{}\phi{}} \right),\\
M_{VU} &= -{\rm{}Im}\left( S_{\phi{}\phi{}} S^*_{\theta{}\theta{}} -
S_{\theta{}\phi{}} S^*_{\phi{}\theta{}} \right),\\
M_{VV} &= {\rm{}Re}\left( S_{\phi{}\phi{}} S^*_{\theta{}\theta{}} -
S_{\theta{}\phi{}} S^*_{\phi{}\theta{}} \right), \label{eq:sca_end}
\end{align}
and
\begin{align}
K_{XX} &= \frac{2\pi{}}{k} {\rm{}Im}\left( S_{\theta{}\theta{}} +
S_{\phi{}\phi{}} \right), \label{eq:ext_start}\\
K_{IQ} &= K_{QI} = \frac{2\pi{}}{k} {\rm{}Im}\left( S_{\theta{}\theta{}} -
S_{\phi{}\phi{}} \right), \\
K_{IU} &= K_{UI} = \frac{2\pi{}}{k} {\rm{}Im}\left( S_{\theta{}\phi{}} +
S_{\phi{}\theta{}} \right), \\
K_{IV} &= K_{VI} = \frac{2\pi{}}{k} {\rm{}Re}\left( S_{\phi{}\theta{}} -
S_{\theta{}\phi{}} \right), \\
K_{QU} &= -K_{UQ} = -\frac{2\pi{}}{k} {\rm{}Im}\left( S_{\phi{}\theta{}} -
S_{\theta{}\phi{}} \right), \\
K_{QV} &= -K_{VQ} = -\frac{2\pi{}}{k} {\rm{}Re}\left( S_{\theta{}\phi{}} +
S_{\phi{}\theta{}} \right), \\
K_{UV} &= -K_{VU} = -\frac{2\pi{}}{k} {\rm{}Re}\left( S_{\phi{}\phi{}} -
S_{\theta{}\theta{}} \right). \label{eq:ext_end}
\end{align}

\subsection{Scattering correction}
\label{sec:correction_convergence}

Once the extinction and scattering matrices are known, the computation of the
absorption coefficients can be done using a numerical integration over all
angles. To this end, we use Gauss-Legendre quadrature on a grid in
$(\cos{}\theta{}, \phi{})$ space (see \sectionref{sec:sampling_grid}). This kind
of numerical quadrature lends itself well for integration over functions that
are smooth functions of direction, and is also extensively used in the T-matrix
calculation itself (see \sectionref{appendix:Tmatrix}).

To quantify the accuracy of our numerical integration scheme, we will compute
the absorption coefficients for a large silicate dust grain ($1~\mu{}$m) and for
the shortest wavelength in our range of interest ($10~\mu{}$m), since these
parameters are the most challenging to compute. For the same reason, we will
focus on prolate and oblate grains at the edges of the integration range for our
assumed shape distribution (see \sectionref{sec:shape_distribution}), i.e.
$d_{\rm{}prol} = 0.19$, $d_{\rm{}obl} = 6.96$. We will then increase the number
of Gauss-Legendre quadrature points and see how the absorption coefficients
converge.

\begin{figure}
\centering{}
\includegraphics[width=0.49\textwidth]{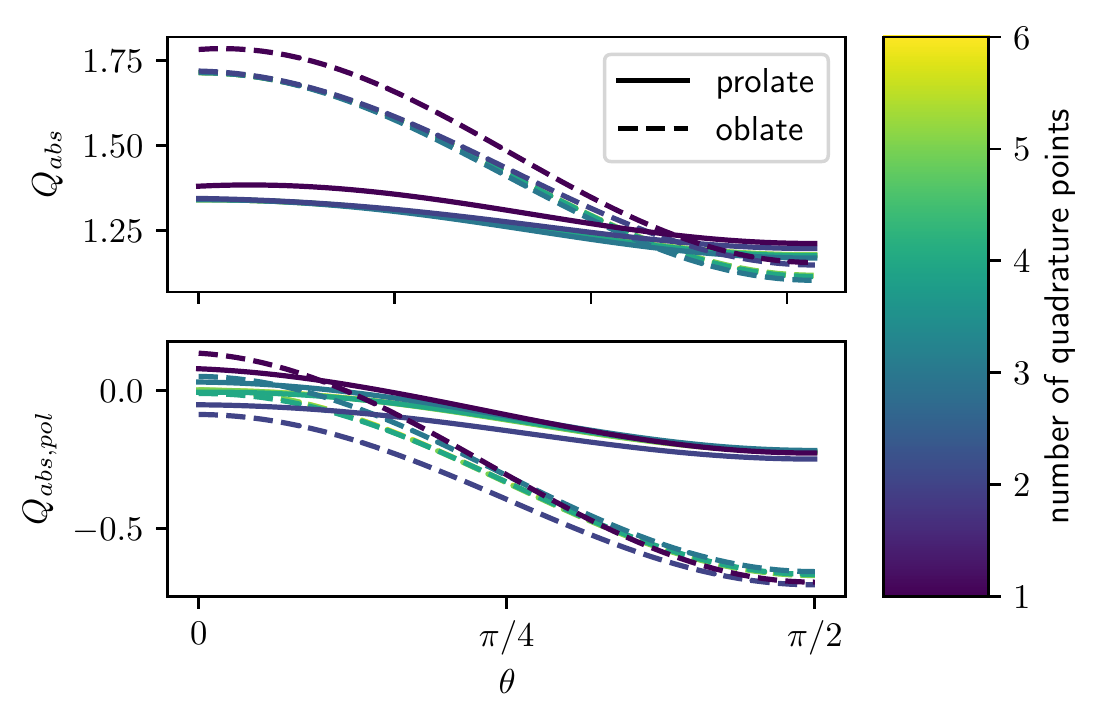}
\includegraphics[width=0.49\textwidth]{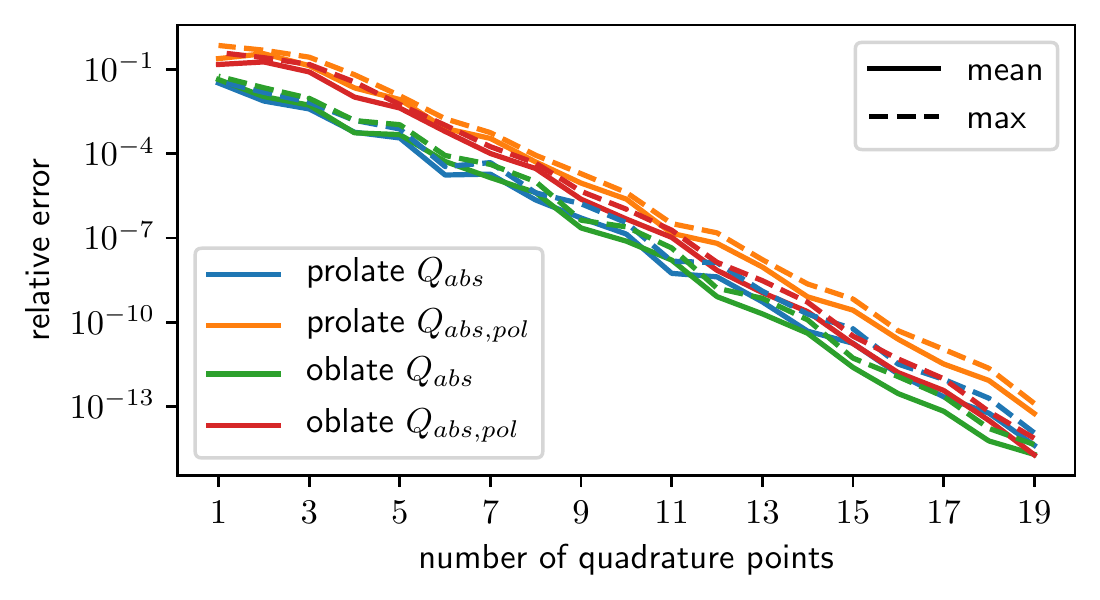}
\caption{\emph{Top:} Convergence of the absorption coefficients with the number
of quadrature points used to compute the scattering correction term.
\emph{Bottom:} Mean and maximum relative error for the same coefficients as a
function of number of quadrature points. The relative error is computed w.r.t.
the mean of the corresponding quantity, using the values obtained with $20^2$
quadrature points as a reference.}
\label{fig:ScatteringCorrection}
\end{figure}

The results of this convergence test are shown in
\figureref{fig:ScatteringCorrection}. The top panel shows the visual convergence
by overplotting results for different numbers of quadrature points (the number
that is quoted corresponds to the number for one of the spherical coordinates;
we sample that number of points in both the azimuth angle $\phi{}$ and the
projected zenith angle $\theta{}$, so the total number of quadrature points
is the square of that number). Due to the smooth nature of the M\"{u}ller matrix
for these input values, convergence is very quick. This is quantified in the
bottom panel, where we show the relative error for the two absorption
coefficients over the full $\theta{}$ range w.r.t. the result obtained using
$20^2$ quadrature points. This corresponds to the value we will use in the rest
of this work.

\subsection{Orientation and alignment}
\label{sec:orientation_and_alignment}

Since \textsc{CosTuuM} is primarily aimed at studying the polarization signature
in thermal emission from aligned dust grains, we have to consider a range of
possible alignment strengths. Note that we make no assumptions about possible
alignment mechanisms: it is the responsibility of the user to choose an
appropriate mechanism for a given grain size, material type and wavelength
within the astrophysical context of interest.

Within our implementation, the alignment of grains is governed by two
independent distributions: an \emph{alignment distribution} that specifies the
alignment strength for grains of different sizes and compositions, and an
\emph{orientation distribution} that expresses the grain alignment for one set
of these parameters as a function of alignment angle. Both of these can be
customized within the \textsc{CosTuuM} framework and we only consider a small
set of examples. For the alignment distribution, we will assume that only
silicate grains with radii larger than a threshold radius of $0.1~\mu{}$m are
aligned, consistent with models that show alignment is small for grains with
sizes smaller than $0.05~\mu{}$m \citep{1995KimMartin, 2009DraineFraisse}, and
with the self-consistently computed alignment thresholds from \citet{2016Reissl}.

Due to symmetry, it is generally possible
to express the orientation of any spheroid w.r.t. the direction of the magnetic
field by means of a single angle $\beta{}$, the angle between the magnetic field
direction and the symmetry axis of the spheroid. The various orientations of the
ensemble of grains can then be captured via a distribution $p(\beta{})$,
normalized such that
\begin{equation}
\int_0^{\pi{}} p(\beta{}) \sin\beta{} {\rm{}d}\beta{} = 1.
\label{eq:orientation_distribution_norm}
\end{equation}

We follow \citet{1991Mishchenko} and focus on three distinct types of
distributions. For grains with a random orientation (no alignment), the
distribution function is uniform and given by
\begin{equation}
p(\beta{})=\frac{1}{2}.
\end{equation}
If the grains are perfectly aligned (this is also called Davis-Greenstein
alignment), the distribution is a Dirac delta distribution:
\begin{align}
p(\beta{}) &= \delta{}\left(\beta{}-\frac{\pi{}}{2} \right), &{\rm{}(prolate)}
\\
p(\beta{}) &= \delta{}\left(\beta{} \right). &{\rm{}(oblate)}
\end{align}
\citet{1991Mishchenko} finally introduces a distribution for imperfectly
aligned grains, given by (note the factor $3$ that is missing in the original
work)
\begin{equation}
p(\beta{}) = \frac{1}{2} + \frac{5}{4}p_2 \left(3\cos^2\beta{}-1\right),
\end{equation}
where $p_2\in{}\left[-1/5,2/5\right]$ is a free parameter. Like
\citet{1991Mishchenko}, we use $p_2=-1/5$ for prolate grains and $p_2=2/5$ for
oblate grains. We will refer to this distribution as the Mishchenko
distribution.

As in \citet{1991Mishchenko} we expand the orientation distribution in Legendre
polynomials and use the expansion coefficients to compute the T-matrix for the
ensemble of grains. This procedure has been implemented in such a way that users
can provide their own expression for the orientation distribution, and we have
successfully tested it with alternative distributions that have the shape of a
normal distribution.

\subsection{Shape distribution}
\label{sec:shape_distribution}

Previous studies have mainly considered a single shape of aligned grains
\citep{2014Siebenmorgen, 2016Reissl, 2017Bertrang}, with a preference for oblate
grains with relatively moderate axis ratios. A real population of grains can be
expected to have a more general distribution $P(d)$ of shapes. Expressions for
these shape distributions can be found in literature \citep{2003Min}, we will
adopt the general Continuous Distribution of Ellipsoids (CDE) given in
\citet{2017Draine}. Their favored CDE, labeled CDE2, is given in terms of the
shape factors $L_i, i=1,2,3$:
\begin{equation}
G(L_1, L_2) = 120 L_1 L_2 L_3 = 120 L_1 L_2 (1 - L_1 - L_2),
\end{equation}
where the shape factors are linked to the axis ratio $d$ through
\begin{equation}
L_1 = L_3 = \frac{1-L_2}{2}
\end{equation}
and
\begin{align}
L_2 &= \frac{1-e^2}{e^2} \left[
  \frac{1}{2e} \ln\left(\frac{1+e}{1-e}\right) -1 \right], \\
e^2 &= 1 - d^2,
\end{align}
for prolate spheroids ($d<1$) and
\begin{align}
L_2 &= \frac{1}{e^2} \left[
  1 - \frac{\sqrt{1-e^2}}{e}\arcsin{}e \right], \\
e^2 &= 1 - \frac{1}{d^2},
\end{align}
for oblate spheroids ($d>1$) \citep{2003Min}.

To obtain a practical expression for $P(d)$ based on the general expression
$G(L_1,L_2)$, we first substitute $L\equiv{}L_2$
\begin{equation}
G(L) = 12 L (1 - L)^2, \label{eq:DHdistribution}
\end{equation}
where the normalization constant was chosen such that $\int_0^1 G(L)
{\rm{}d}L=1$. For each value of $d\in{}[0, \infty{}[$, we can determine the
shape factor $L$ and the corresponding CDE2 probability density. To turn this
into a probability density function for the coordinate $d$, we need to multiply
with the Jacobian \citep{2003Min}
\begin{equation}
\frac{{\rm{}d}L}{{\rm{}d}d} = \frac{\sqrt{1-e^2}}{2e^5} \left[
  (3-e^2) \ln\left(\frac{1+e}{1-e}\right) -6e \right]
\label{eq:DHjacprol}
\end{equation}
for prolate spheroids and
\begin{equation}
\frac{{\rm{}d}L}{{\rm{}d}d} = \frac{1-e^2}{e^5} \left[
  (3-2e^2) \arcsin{}e - 3e\sqrt{1-e^2} \right]
\label{eq:DHjacobl}
\end{equation}
for oblate spheroids.

\begin{figure}
\centering{}
\includegraphics[width=0.5\textwidth]{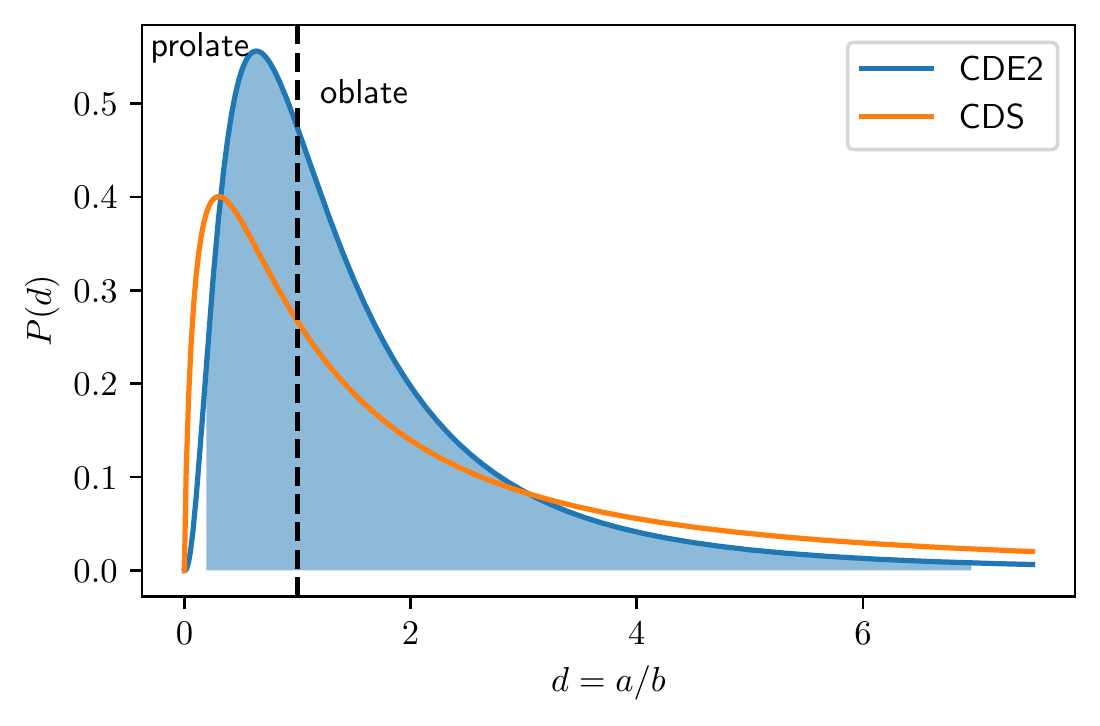}
\caption{Shape distribution as a function of axis ratio, $d$. Shown are the
\citet{2017Draine} CDE2 distribution and the \citet{2003Min} CDS distribution.
The shaded area underneath the CDE2 curve corresponds to the 96~\% of the
distribution that we actually sample in our runs.}
\label{figure:shape_distribution}
\end{figure}

The corresponding distribution function as a function of the axis ratio $d$ is
shown in \figureref{figure:shape_distribution}. Also shown is the continuous
distribution of spheroids (CDS) expression given in \citet{2003Min} that assumes
a uniform weight for all values of $L$. While in the latter, $1/3$ of the
grains are prolate, the CDE2 distribution has relatively less extreme axis
ratios, so that $\approx{}40$~\% of the grains is prolate.

As for the scattering correction, we will numerically average the absorption
coefficients over the shape distribution using Gauss-Legendre quadrature.
This procedure is novel for our method, and hence no comparison results are
available in literature. We can check the accuracy of our implementation in two
steps. First, we need to show that our numerical quadrature rule is implemented
correctly. This can be done by replacing the input absorption coefficients and
the assumed shape distribution with known analytic functions, and checking
against the expected analytic result. This is a trivial exercise. Second, we
need to check that our results are converged for the choice of parameters we
make for the numerical quadrature. This can again be achieved by increasing the
number of quadrature points until convergence is reached.

\begin{figure}
\centering{}
\includegraphics[width=0.49\textwidth]{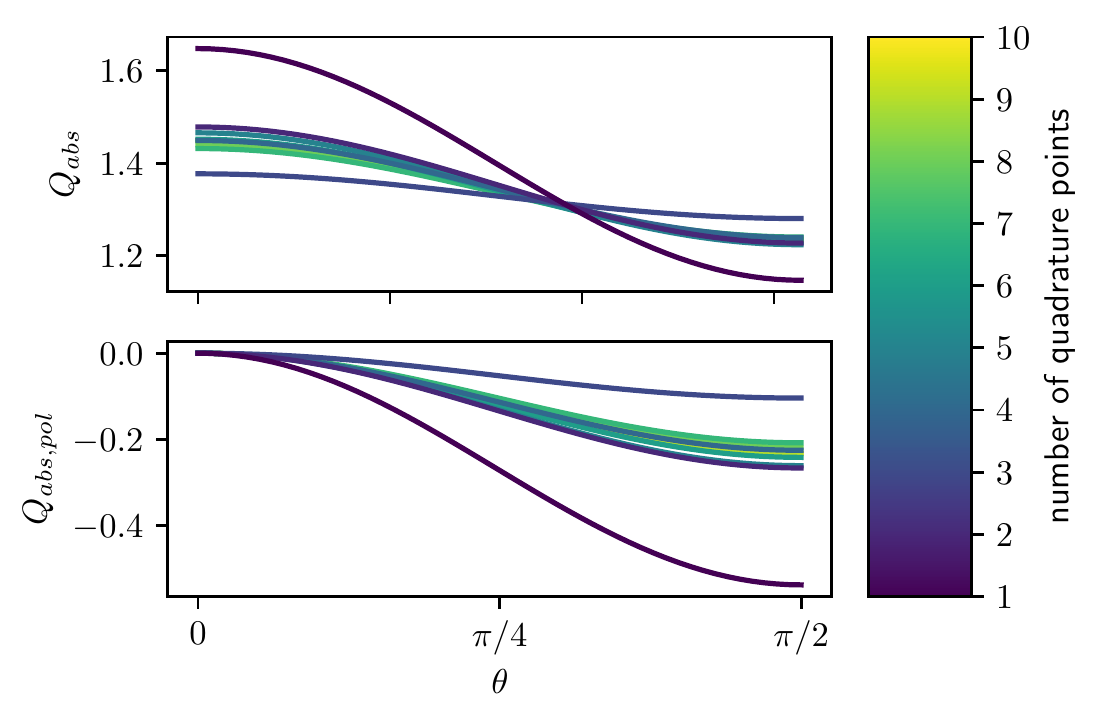}
\includegraphics[width=0.49\textwidth]{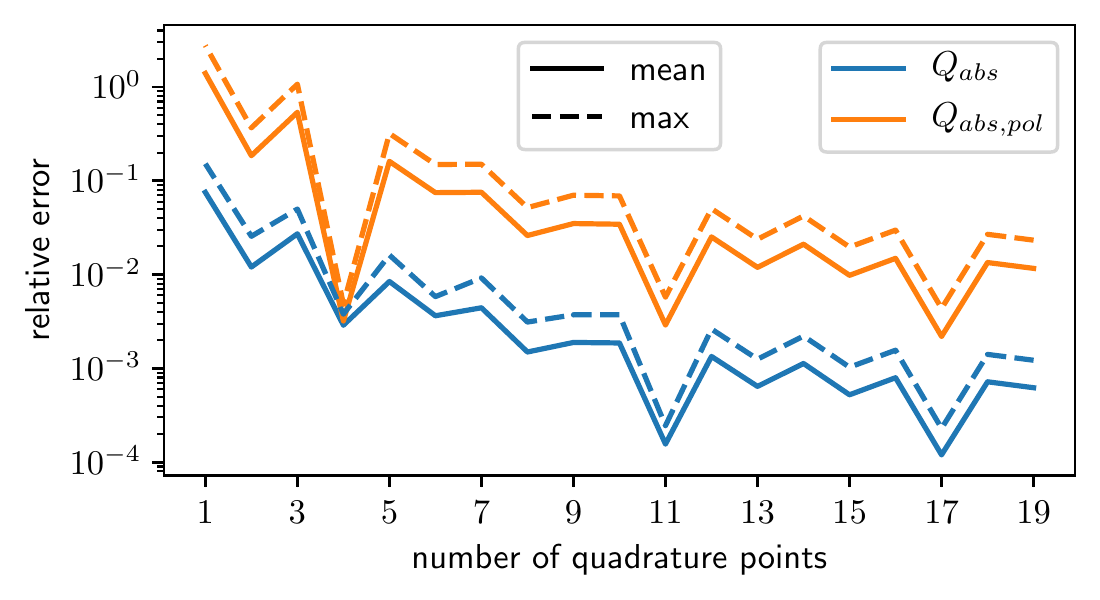}
\caption{\emph{Top:} Convergence of the absorption coefficients with the number
of quadrature points used to compute the shape distribution average.
\emph{Bottom:} Mean and maximum relative error for the same coefficients as a
function of number of quadrature points. The relative error is computed w.r.t.
the mean of the corresponding quantity, using the values obtained with 20
quadrature points as a reference.}
\label{fig:ShapeQuadrature}
\end{figure}

\figureref{fig:ShapeQuadrature} shows the convergence results for this test.
Since in this case every sample in shape space corresponds to a separate
T-matrix calculation, the absolute error is quickly dominated by the error on
a single T-matrix calculation. For the absorption coefficient, we can achieve
a relative error of $\approx{}10^{-3}$. For the polarized absorption coefficient
we achieve a similar absolute error, but since the actual values are lower, we
are limited to a relative error of a few percent.

\begin{figure}
\centering{}
\includegraphics[width=0.49\textwidth]{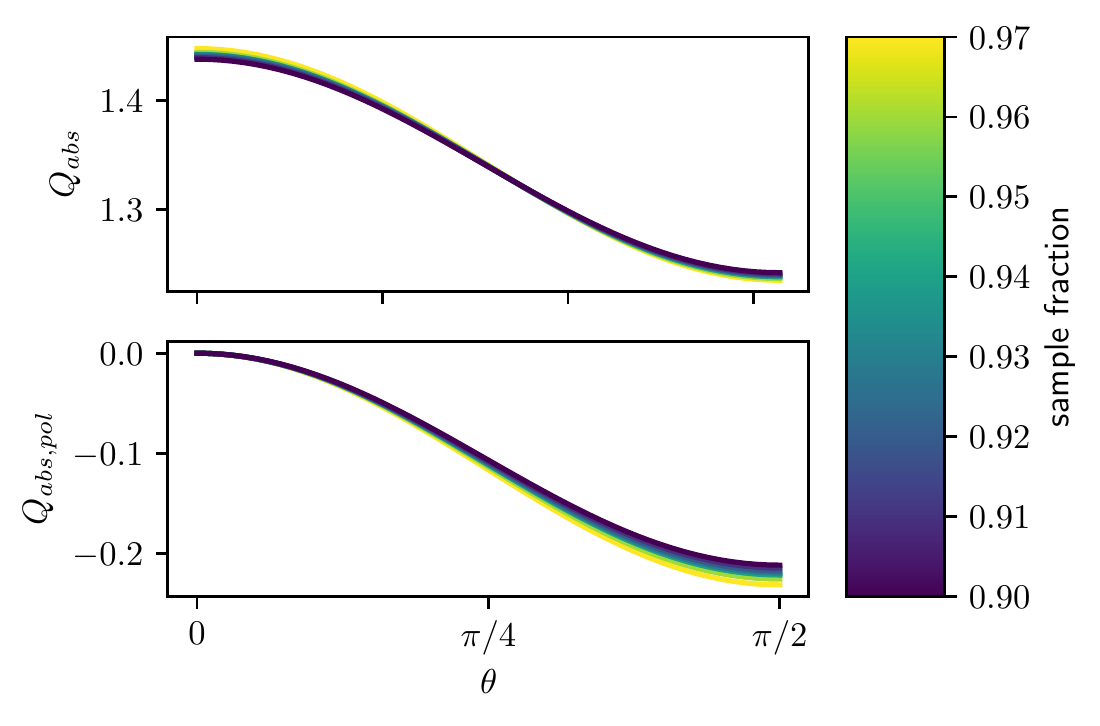}
\includegraphics[width=0.49\textwidth]{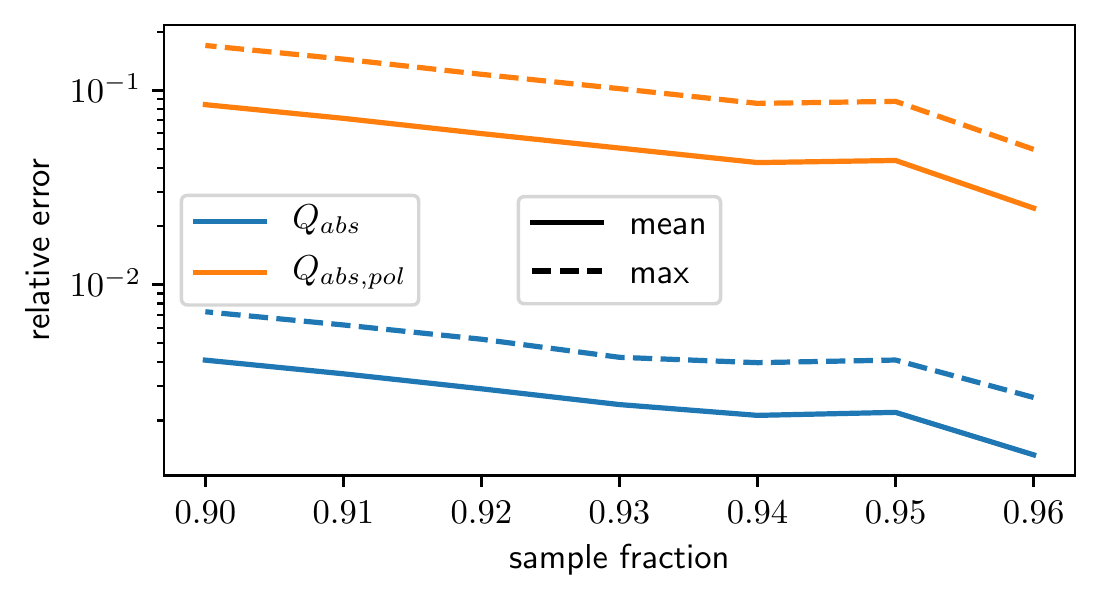}
\caption{\emph{Top:} Convergence of the absorption coefficients with the
fraction of the shape distribution that gets sampled.
\emph{Bottom:} Mean and maximum relative error for the same coefficients as a
function of sampling fraction. The relative error is computed w.r.t. the mean of
the corresponding quantity, using the values computed with a sampling fraction
$f_S=0.97$ as a reference.}
\label{fig:ShapeBounds}
\end{figure}

Another issue to address is the fraction of the shape distribution that we need
to sample to get consistent absorption coefficients. In principle, the shape
distribution is defined for axis ratio values $d\in{}[0, \infty{}[$, but the
T-matrix method runs into numerical issues for axis ratio values close to both
ends of that interval. We therefore determine new limits, $[d_l, d_u]$ so that
we sample a representative fraction $f_S$ of the CDE2 distribution on both ends
of the axis $d=1$:
\begin{equation}
f_S = \frac{\int_{d_l}^1 P(d) {\rm{}d}d}{\int_0^1 P(d) {\rm{}d}d}
= \frac{\int_1^{d_u} P(d) {\rm{}d}d}{\int_1^\infty{} P(d) {\rm{}d}d}
\end{equation}
and assume the entire distribution can be approximated by this value.

\figureref{fig:ShapeBounds} shows the absorption coefficients for various
fractions of the sampling interval, for the same challenging grain size and
wavelength as used above. The method breaks down for fractions larger than
97~\%. It is clear that our results are still dependent on the sampling fraction
for that value, but the dependence is only of the order of 10~\% for the most
extreme zenith angles, and overall the results are reasonably converged. We
therefore settle on a value of 96~\% for all our calculations. Note that the
numerical issues reduce the fraction we can sample quickly once $\lambda{}$
decreases: for a grain size of $a=1~\mu{}$m, we can only sample 80~\% of the
distribution at $\lambda{}=1~\mu{}$m, and this even goes down to 50~\% at
$\lambda{}=0.3~\mu{}$m. As shown by \citet{2013Somerville}, these
issues can be alleviated by making some changes to the core of the T-matrix
algorithm, should this be required in the future.

\subsection{Grain properties}

Apart from the alignment and shape properties already mentioned above, we also
need to specify the material properties (refractive index) and size distribution
for the dust grains. For the former, we are limited to tabulated values
available in literature, e.g. \citet{1985Draine}. Alternatively, we also made
it possible to pass on custom material properties to the CosTuuM library.

Since grain sizes are a function of position in many contemporary RT methods, we
do not integrate the optical properties over a grain size distribution within
CosTuuM and leave this to the application. However, to illustrate the impact of
a grain mixture of different sizes on the results shown in
\sectionref{sec:results}, we will assume a basic MRN power-law grain size
distribution \citep{1977Mathis}:
\begin{equation}
\Omega{}(a) \propto{} a^{-3.5}, a \in{} [a_{\rm{}min}, a_{\rm{}max}].
\end{equation}
Following \citet{2016Reissl}, we set $a_{\rm{}min}=5$~nm and
$a_{\rm{}max}=2~\mu{}$m. This additional grain size averaging procedure is not
part of \textsc{CosTuuM}, and was entirely implemented using Python's NumPy
library.

\section{Results and discussion}
\label{sec:results}

In this section, we will apply \textsc{CosTuuM} to a range of scenarios to
illustrate its validity and accuracy, and to explore the impact of some commonly
made assumptions about the properties of spheroidal dust grains.

\subsection{Mishchenko extinction results}
\label{sec:Mishchenko_benchmark}

\citet{1991Mishchenko} contains a small number of reference values for the
extinction coefficients, computed using the same T-matrix formalism used by
\textsc{CosTuuM}. Reproducing these values correctly is a good test for the
T-matrix formalism itself and for the procedure to average over different
particle orientations (a part not available in the public version of
Mishchenko's code).

The test assumes a single silicate dust grain with a radius of $0.2~\mu{}$m that
is either prolate with an axis ratio $d=1/2$ or oblate with an axis ratio $d=2$.
For both grains, the three different alignment distributions are compared:
random orientation (no alignment), perfect (Davis-Greenstein) alignment, and
Mishchenko alignment.

For perfect and imperfect alignment, the extinction coefficients become
dependent on the angle between the direction of the magnetic field and the
incoming electromagnetic wave, so values are listed for 4 different angles:
$\theta{}=\{0, \pi{}/6, \pi{}/3, \pi{}/2\}$. The refractive index of the grain
is set to the values from \citet{1985Draine}, which differ slightly from the
values given on Bruce Draine's web page that are also used in
\sectionref{sec:draine}.

\begin{figure*}
\centering{}
\includegraphics[width=0.98\textwidth]{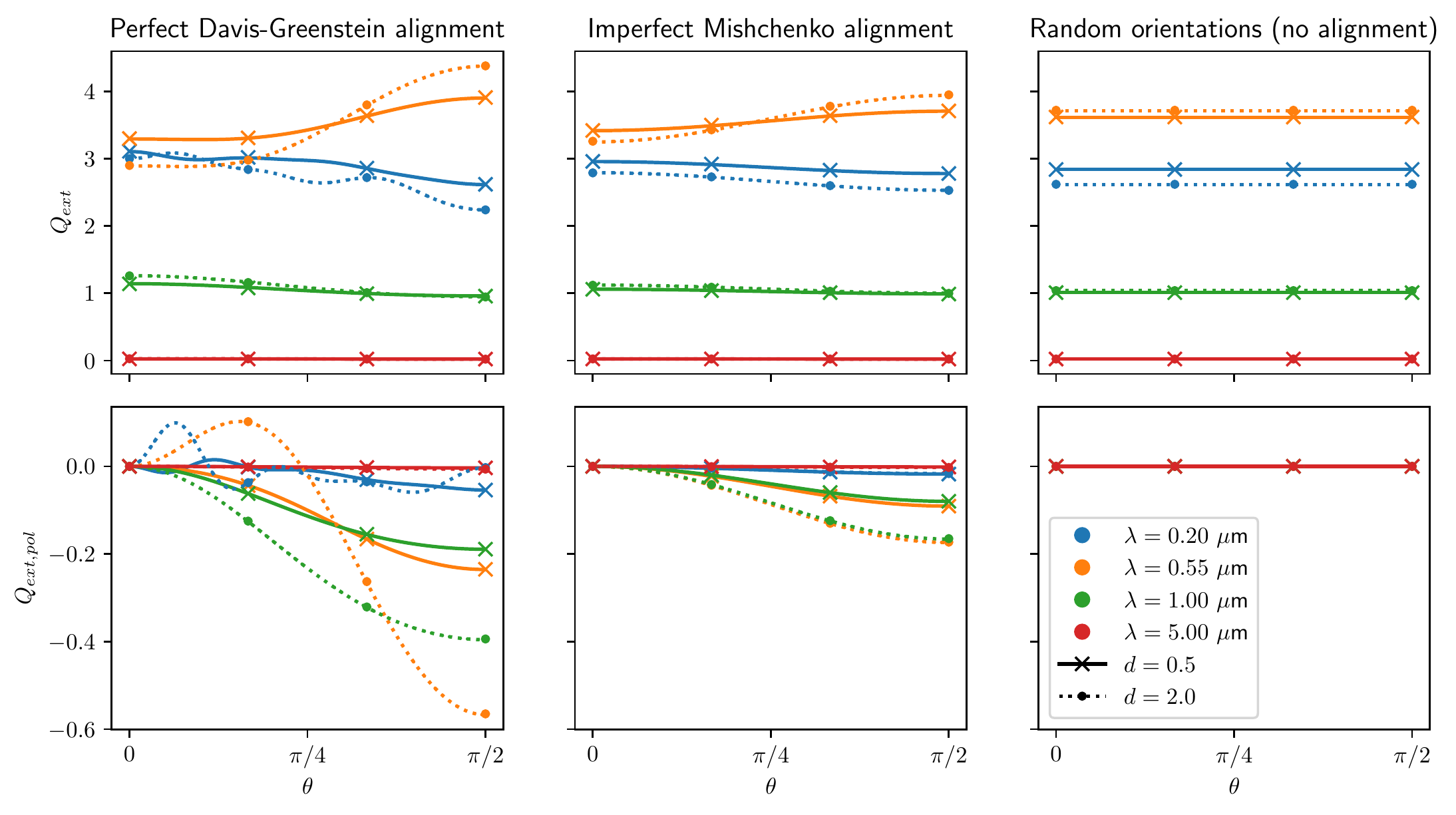}
\caption{Extinction coefficient (\emph{top row}) and linearly polarized
extinction coefficient (\emph{bottom row}) as a function of zenith angle
$\theta{}$, for a silicate dust grain with an equal volume radius $0.2~\mu{}$m,
and for four different incoming wavelengths and two different axis ratios, as
indicated in the legend. The three columns correspond to three different
alignment distributions. The symbols correspond to the values from
\citet{1991Mishchenko}, while the lines were computed with \textsc{CosTuuM} on a
linear $\theta{}$ grid with 100 values.}
\label{fig:1991Mishchenko}
\end{figure*}

The results of the tests are shown in \figureref{fig:1991Mishchenko}. The
results are in clear agreement for all wavelengths and both axis ratios, and for
all alignment cases. Due to the significant improvements in computational power,
we can now very easily compute these values for many more angles than in the
original work by \citet{1991Mishchenko}.

For long wavelengths and small dust grains (or small values of the
size parameter $x_V=2\pi{}a/\lambda{}$, commonly used in literature),
scattering by dust becomes negligible, so the extinction coefficients in this
regime will be equal to the absorption coefficients. In this regime, this test
hence directly tests the validity of the absorption coefficients. However, for
shorter wavelengths, this approximation no longer holds and we need to also take
into account the scattering correction term in equation \eqref{eq:absorption}.
We need to test this part of the algorithm separately.

\subsection{Voshchinnikov extinction results}

\citet{1993Voshchinnikov} present another set of reference results for the
extinction coefficients, this time computed using the SVM method. Their results
do not include any orientation averaging, but provide an independent benchmark
test for the results of an individual scattering calculation. We present three
different tests, respectively corresponding to Figs. 3-5 in
\citet{1993Voshchinnikov}. We compare \textsc{CosTuuM} results with results
obtained using the public version of the \citet{1993Voshchinnikov} code. All
these tests use a fixed photon wavelength $\lambda{}=0.5~\mu{}$m.

\begin{figure*}
\centering{}
\includegraphics[width=0.98\textwidth]{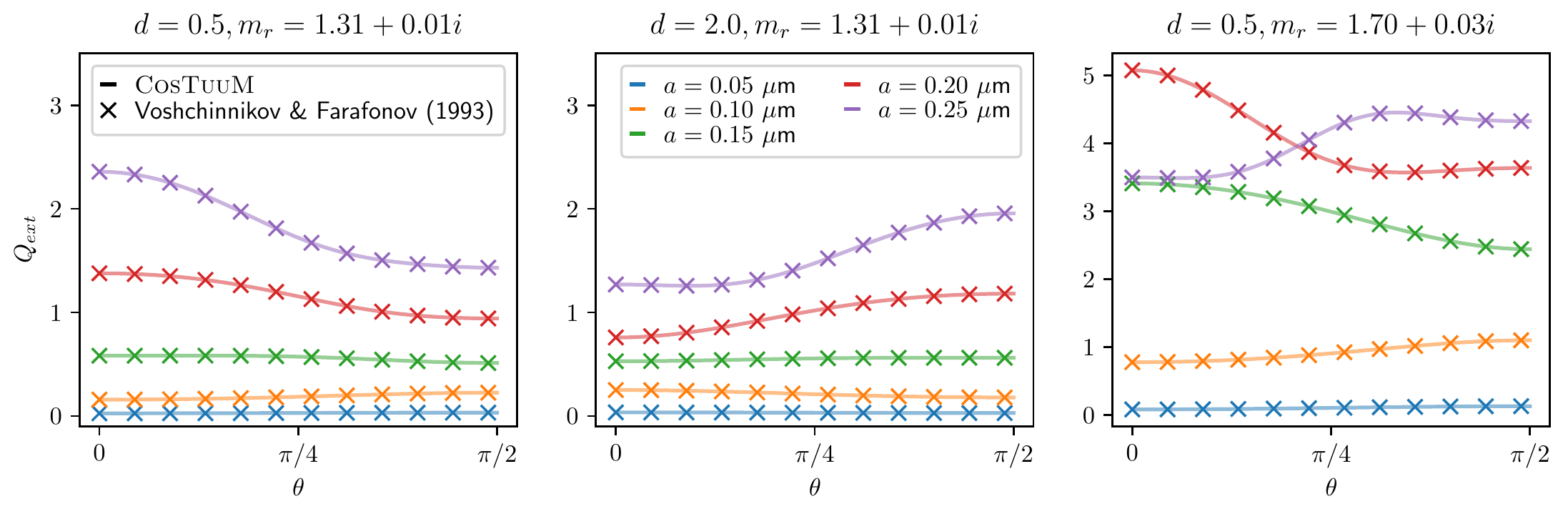}
\caption{Extinction coefficient as a function of zenith angle $\theta{}$, for a
single spheroidal dust grain that has its symmetry axis oriented along the
positive $z$ axis. The shape and refractive index of the grain are indicated
above each panel, while the grain size is indicated in the legend. Full lines
correspond to results computed with \textsc{CosTuuM}, while the symbols are the
results from \citet{1993Voshchinnikov}. All results use a photon wavelength
$\lambda=0.5~\mu{}$m.}
\label{fig:VoshchinnikovFig3}
\end{figure*}

The first test evaluates the zenith angle dependence of the
extinction coefficient $Q_{ext}$ for five different grain sizes ($a\in{}[0.05,0.10,0.15,0.20,0.25]~\mu{}$m), and for moderate axis
ratios, $d=1/2$ and $d=2$. Results are obtained for prolate grains with
refractive index $m_r=1.31+0.01i$ and $m_r=1.70+0.03i$, while only the former
value is used for the oblate spheroid. As shown in
\figureref{fig:VoshchinnikovFig3}, our results are in excellent agreement.

\begin{figure*}
\centering{}
\includegraphics[width=0.98\textwidth]{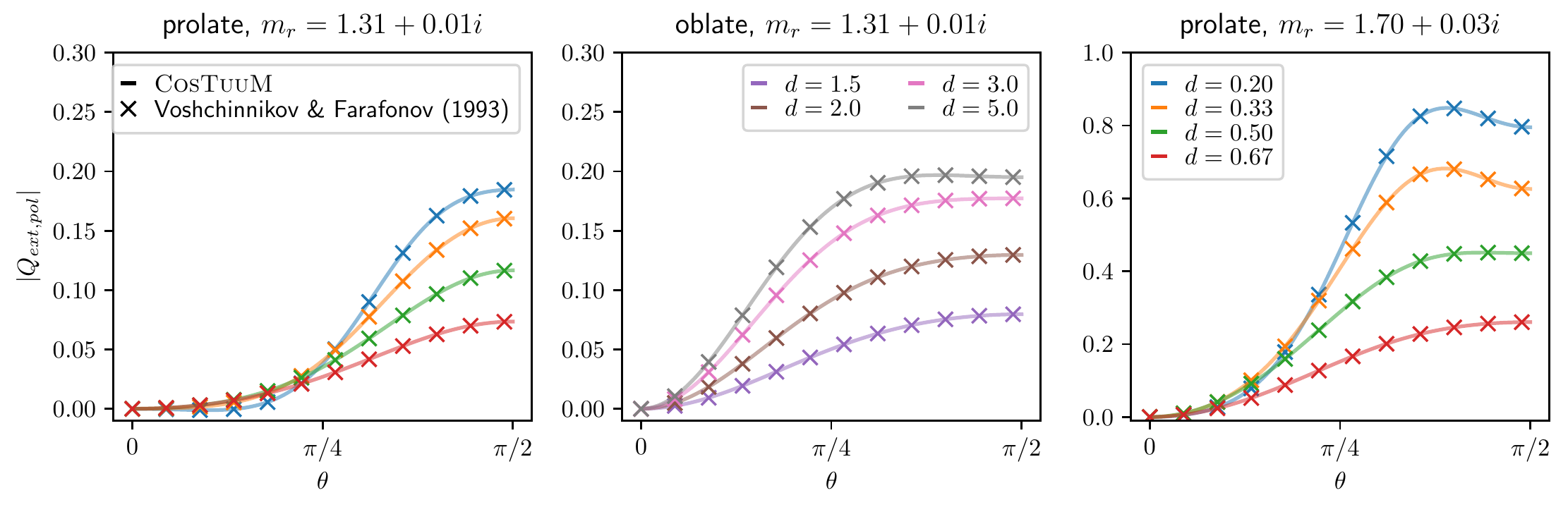}
\caption{Polarized extinction coefficient as a function of zenith angle
$\theta{}$, for a single spheroidal dust grain that has its symmetry axis
oriented along the positive $z$ axis. The refractive index of the grain is
indicated above each panel, while the shape is indicated in the legend. Full
lines correspond to results computed with \textsc{CosTuuM}, while the symbols
are the results from \citet{1993Voshchinnikov}. All results use a grain size
$a=0.25~\mu{}$m and photon wavelength $\lambda{}=0.5~\mu{}$m.}
\label{fig:VoshchinnikovFig4}
\end{figure*}

The second test looks at the zenith angle dependence of the polarized exinction
coefficient $Q_{ext, pol}$ for different axis ratios $d$, and uses a single
grain size $a=0.25~\mu{}$m and refractive indices $m_r=1.31+0.01i$ (for both
prolate and oblate grains) and $m_r=1.70+0.03i$ (prolate grains only).
\figureref{fig:VoshchinnikovFig4} again shows excellent agreement.

\begin{figure}
\centering{}
\includegraphics[width=0.48\textwidth]{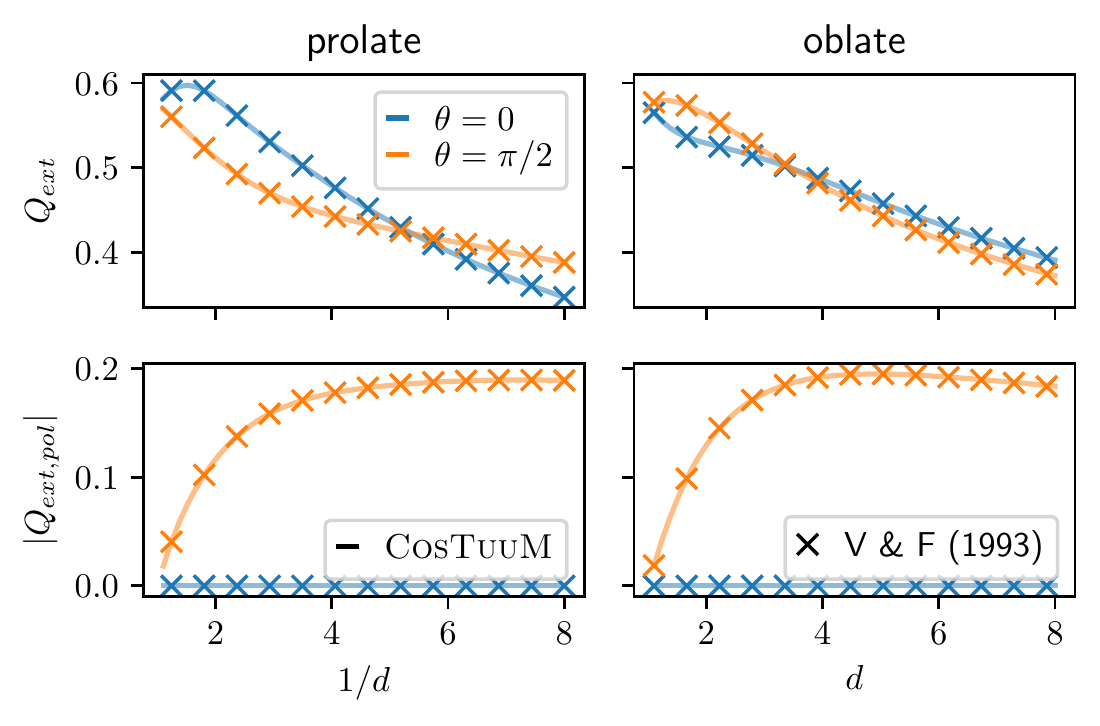}
\caption{Extinction coefficient (\emph{top}) and polarized extinction
coefficient (\emph{bottom}) as a function of (inverse) axis ratio $d$, for a
single spheroidal dust grain that has its symmetry axis oriented along the
positive $z$ axis. Results are shown for two zenith angles $\theta{}$, as
indicated in the legend. Full lines correspond to results computed with
\textsc{CosTuuM}, while the symbols are the results from
\citet{1993Voshchinnikov}. All results use a refractive index $m_r=1.31+0.01i$,
grain size $a=0.25~\mu{}$m and photon wavelength $\lambda{}=0.5~\mu{}$m. To
facilitate comparison with \citet{1993Voshchinnikov}, the inverse axis ratio is
plotted for the prolate
grains.}
\label{fig:VoshchinnikovFig5}
\end{figure}

The final test considers the axis ratio dependence of both $Q_{ext}$ and
$Q_{ext,pol}$ at zenith angles $\theta{}=0$ and $\theta{}=\pi{}/2$, for a single
grain with size $a=0.25~\mu{}$m and refractive index $m_r=1.31+0.0i$. As seen in
\figureref{fig:VoshchinnikovFig5}, we obtain excellent agreement over the full
range $d\in{}[0.13,8.00]$.

\subsection{Draine dust models}
\label{sec:draine}

To test the calculation of absorption coefficients (including the scattering
correction), we compare with the publicly available values of
\citet{1984Draine,1993Draine}. They provide values for a large wavelength range
and for a number of grain sizes, but only for spherical grains.

\begin{figure}
\centering{}
\includegraphics[width=0.49\textwidth]{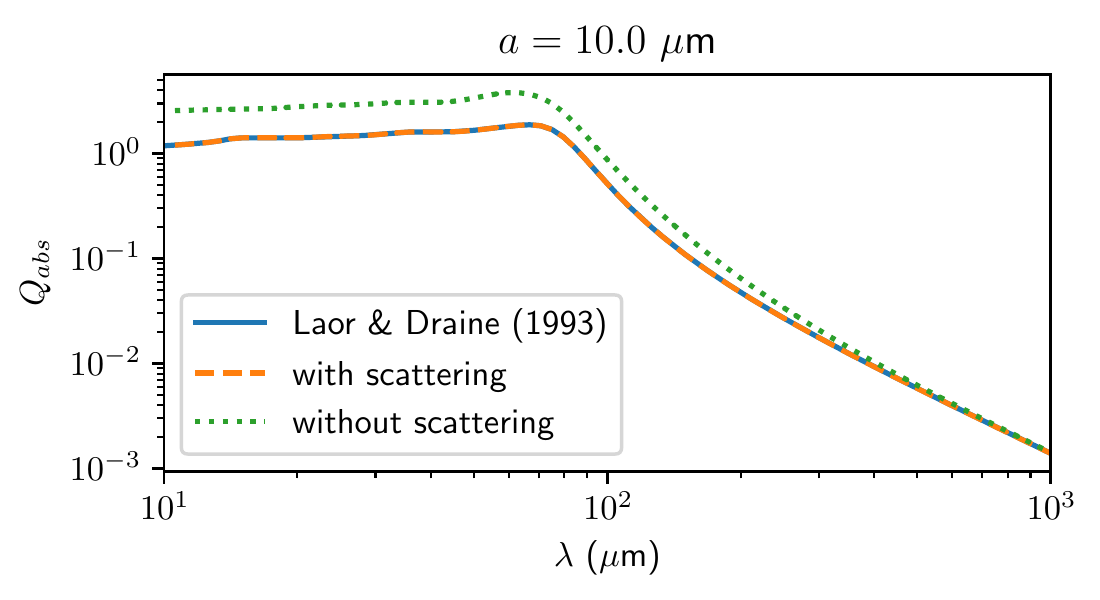}
\caption{Absorption coefficient as a function of wavelength for a single
spherical silicate dust grain. The different lines compare the original
\citet{1993Draine} with values calculated using \textsc{CosTuuM}, with and
without the additional scattering correction term.}
\label{fig:1993Draine_spherical}
\end{figure}

\figureref{fig:1993Draine_spherical} shows an example for one grain size for a
limited wavelength range, and shows the impact of including the scattering
correction, especially at shorter wavelengths. Note that the T-matrix method
becomes unstable for short wavelengths and large grain sizes, so that our
\textsc{CosTuuM} calculations are necessarily limited to this range. This is not
an issue, since thermal dust emission becomes negligible for short wavelengths.

\begin{figure}
\centering{}
\includegraphics[width=0.49\textwidth]{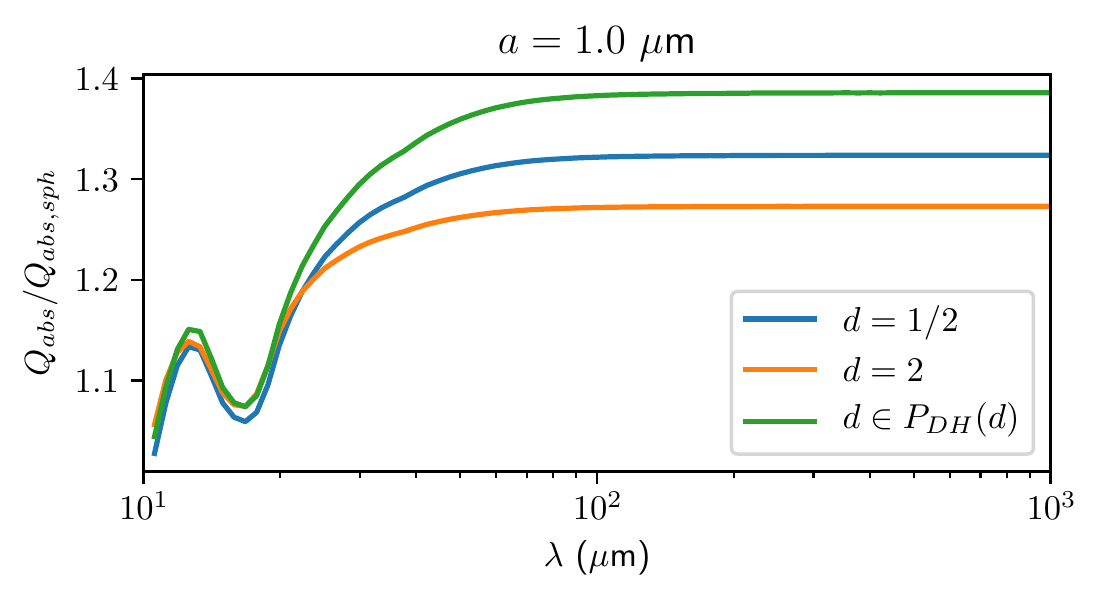}
\caption{Ratio of the absorption coefficient for spheroidal grains and for
spherical grains with the same equal volume radius, as a function of wavelength,
and for three different shapes, as indicated in the legend.}
\label{fig:1993Draine_shapes}
\end{figure}

\figureref{fig:1993Draine_shapes} shows the ratio of the absorption coefficients
for spheroidal grains and for spherical grains with the same equal volume
radius, for a pure oblate ($d=2$) and pure prolate ($d=1/2$) silicate grain and
for a realistic mixture of grains with the shape distribution given by equations
\eqref{eq:DHdistribution}-\eqref{eq:DHjacobl}. Note that the ratios are always
larger than one, indicating that spheroidal grains are more efficient absorbers
than the equivalent spherical grains. Moreover, the grains with shapes
distributed according to the CDE2 distribution have significantly higher
absorption coefficients, since their average value is dominated by oblate
spheroids with relatively large axis ratios.

One explanation for this increased efficiency is that the definition of the
absorption/extinction coefficient given by equations
\eqref{eq:Qext}-\eqref{eq:Qabs} ignores the fact that spheroidal grains have a
larger active surface area, i.e. the total average projected surface area seen
by photons that are incoming from all sides is always larger for spheroidal
grains than for a sphere with the same volume. As a result, the absorption
coefficients for both oblate and prolate spheroidal grains are always larger
than for a spherical grain with the same volume. We can correct for this effect
by multiplying with the corresponding correction factor,
\begin{equation}
C_A = \left( 
  \frac{d^{-1/3}}{2} \int_{-1}^{1} \sqrt{
    \frac{(d^4-1)x^2 + 1}{(d^2-1)x^2 + 1}} dx
\right)^{-1}.
\end{equation}
This expression can be derived using basic geometrical algebra
\citep{2001Vickers}.

\begin{figure}
\centering{}
\includegraphics[width=0.49\textwidth]{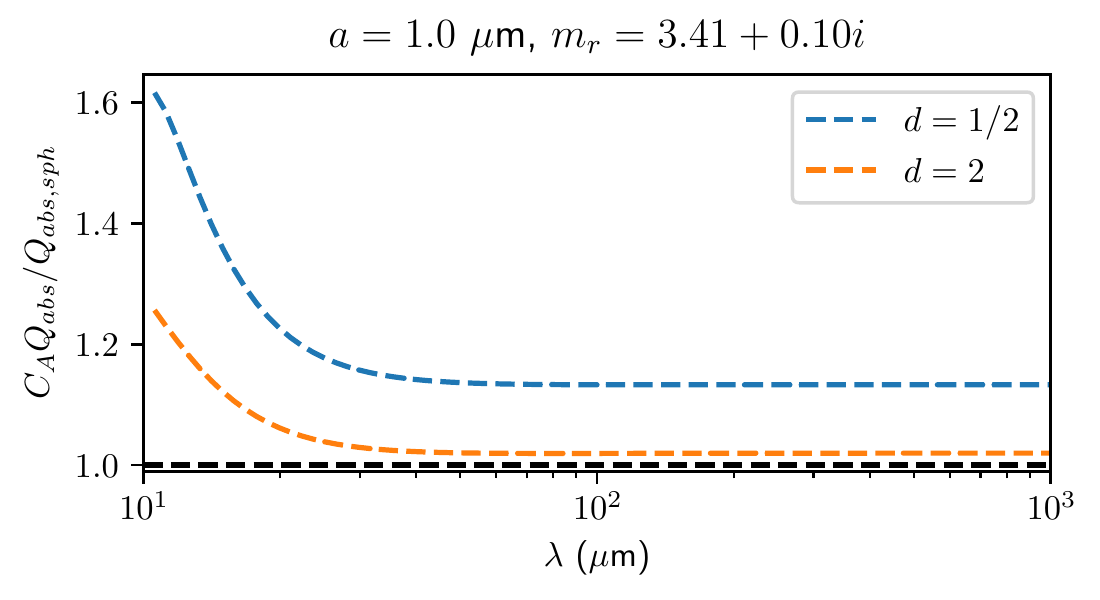}
\caption{Ratio of the absorption coefficient for spheroidal grains and for
spherical grains with the same equal volume radius, as a function of wavelength,
corrected to account for the higher average projected surface area. To eliminate
additional differences because of the size dependence of the refractive index,
we assume a constant refractive index.}
\label{fig:1993Draine_correction}
\end{figure}

To test that this is indeed the case, we need to eliminate all other differences
in the calculation for the absorption coefficient, i.e. we need to make sure we
use a constant refractive index. \figureref{fig:1993Draine_correction} shows the
results for a refractive index $m_r=3.41+0.10 i$ (corresponding to the
\citet{1993Draine} value for $\lambda{}=500~\mu{}$m), this time corrected with
the $C_A$ factor. The corrected absorption coefficients get closer to the values
for the equivalent spherical grains, although they are still higher; this is a
manifestation of the \emph{extinction paradox} \citep{1983BohrenHuffman}. The
offset gets larger when the wavelength of the incoming radiation gets closer to
the size of the particle due to non-linear effects.

For an assumed constant grain composition, the mass of the dust grain is
proportional to its volume. So the different absorption coefficients can be
readily compared if we assume a constant dust mass. As can be seen from
\figureref{fig:1993Draine_shapes}, dust absorption and emission are stronger for
spheroidal dust grains.

\begin{figure}
\centering{}
\includegraphics[width=0.49\textwidth]{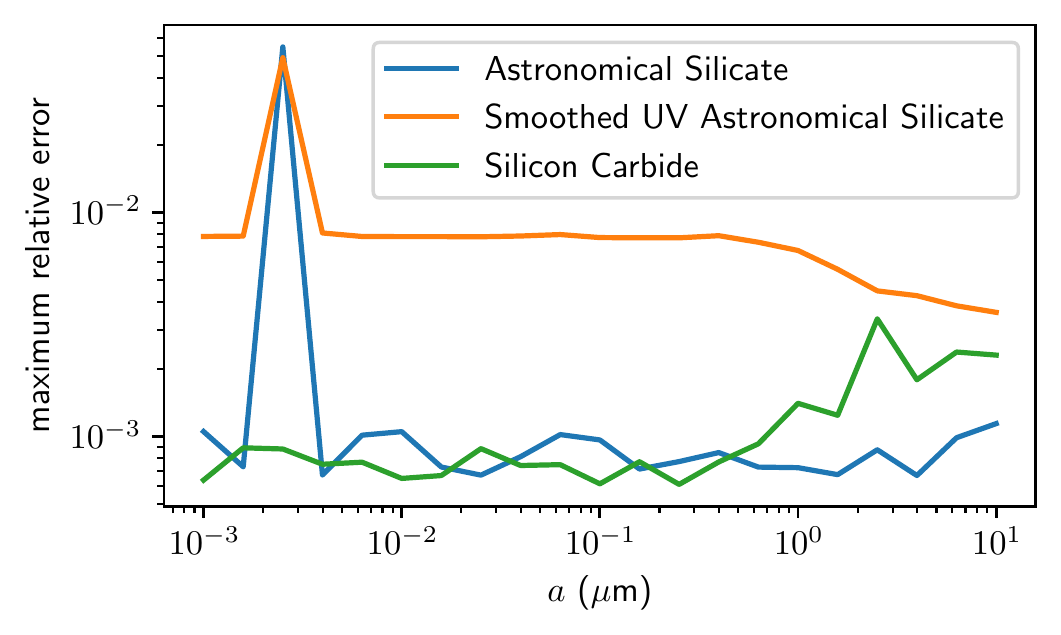}
\caption{Maximum relative difference between \textsc{CosTuuM} absorption
coefficients and the values presented by \citet{1984Draine, 1993Draine} as a
function of grain size, for three different grain types available on Bruce
Draine's web page, as indicated in the legend.}
\label{fig:1993Draine_all}
\end{figure}

\begin{figure}
\centering{}
\includegraphics[width=0.49\textwidth]{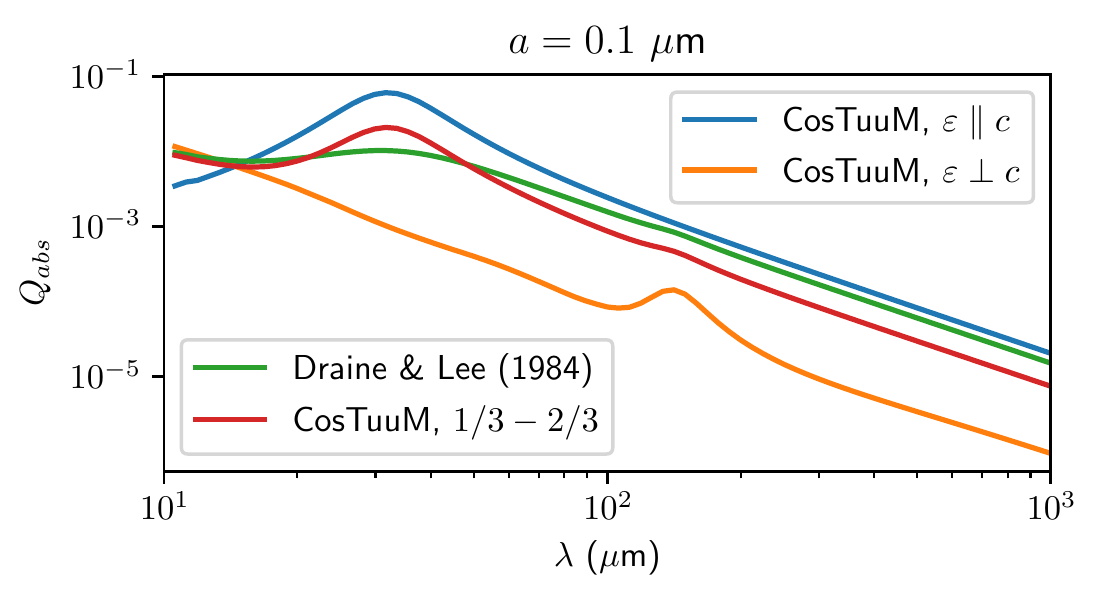}
\caption{Absorption coefficient as a function of wavelength for carbon dust
grains with the size indicated in the title. The different colors correspond to
different \textsc{CosTuuM} models and the reference values from
\citet{1984Draine}, as indicated in the legend. The $1/3-2/3$ \textsc{CosTuuM}
model corresponds to the weighted average of the $\varepsilon{}\parallel{}c$ and
$\varepsilon{}\perp{}c$ models, whereby the former have $1/3$ of the weight and
the latter the other $2/3$.}
\label{fig:1993Draine_carbon}
\end{figure}

\figureref{fig:1993Draine_all} shows the relative difference between the
\textsc{CosTuuM} absorption coefficients and the reference models computed by
\citet{1984Draine, 1993Draine}, as a function of grain size and for various
grain compositions. Overall, the relative error is low, so that we can conclude
that \textsc{CosTuuM} reproduces the existing optical properties results very
well. A notable exception is carbon (see \figureref{fig:1993Draine_carbon}), for
which \textsc{CosTuuM} systematically under predicts the absorption coefficient
for long wavelengths and over predicts some features at shorter wavelengths. As
pointed out by \citet{1984Draine}, carbon is a highly anisotropic material, so
that its dielectric function is in fact a tensor; this tensor is usually
represented in its diagonal form, with one diagonal element parallel to the
incident electromagnetic field ($\varepsilon{}\parallel{}c$), and two elements
perpendicular to it ($\varepsilon{}\perp{}c$). The optical properties of carbon
are then computed by taking an ad-hoc $1/3$-$2/3$ mixture of the optical
properties computed for both values of the refractive index. On top of that,
carbon is a dielectric material with a non-negligible magnetic dipole
absorption, which is not treated accurately by the T-matrix method.

\subsection{Polarized dust emission}

Now that we have shown that \textsc{CosTuuM} generates accurate results in the
infrared wavelength regime, we can use it to make predictions for the absorption
cross sections for a reference silicate grain model with an MRN size
distribution, $\Omega{}(a)=a^{-3.5}$, a CDE2 shape distribution, and perfect
grain alignment. We will compute the size-averaged absorption cross sections
\begin{align}
\langle{} K_{a,I} \rangle{}(\lambda{}, \theta{}) &=
\int_{a_{\rm{}min}}^{a_{\rm{}max}} \Omega{}(a)
Q_{\rm{}abs}(\lambda{},a,\theta{}) \pi{} a^2 {\rm{}d}a, \\
\langle{} K_{a,Q} \rangle{}(\lambda{}, \theta{}) &=
\int_{a_{\rm{}min}}^{a_{\rm{}max}} \Omega{}(a)
Q_{\rm{}abs,pol}(\lambda{},a,\theta{}) \pi{} a^2 {\rm{}d}a.
\end{align}
We will evaluate these functions at three representative wavelengths:
$70~\mu{}$m, $200~\mu{}$m and $350~\mu{}$m. These correspond to the peak
wavelengths for black-body emission at $72.9$~K, $25.5$~K and $14.6$~K, and are
also the centers of the bands for the B-BOP polarimeter on board the proposed
ESA/JAXA M5 mission SPICA \citep{2018Roelfsema,2019Andre}. We will only show the
zenith angle range $[0, \pi{}/2]$, since our results are symmetric w.r.t. the
$\theta{}=\pi{}/2$ axis. Independent of the assumed alignment mechanism, all our
results assume a magnetic field directed along the positive $z$ axis with
$\theta{}=0$.

As in \citet{2017Peest}, we will characterize the strength of the polarization
using the degree of linear polarization,
\begin{equation}
P_L = \frac{\sqrt{Q^2 + U^2}}{I} = \frac{\left| \langle{} K_{a,Q}
\rangle{}(\lambda{}, \theta{}) \right|}{\langle{} K_{a,I} \rangle{}(\lambda{},
\theta{})},
\end{equation}
where in the second step we have assumed that the observed intensities are
directly proportional to the emitted intensities and are observed with an
instrument that has its north direction parallel to the orientation of the
magnetic field, so that $U=0$. The linear polarization angle
\begin{equation}
\gamma{} = \frac{1}{2} \arctan\left(\frac{U}{Q}\right)
\end{equation}
is in this case trivially 0.

\begin{figure}
\centering{}
\includegraphics[width=0.48\textwidth]{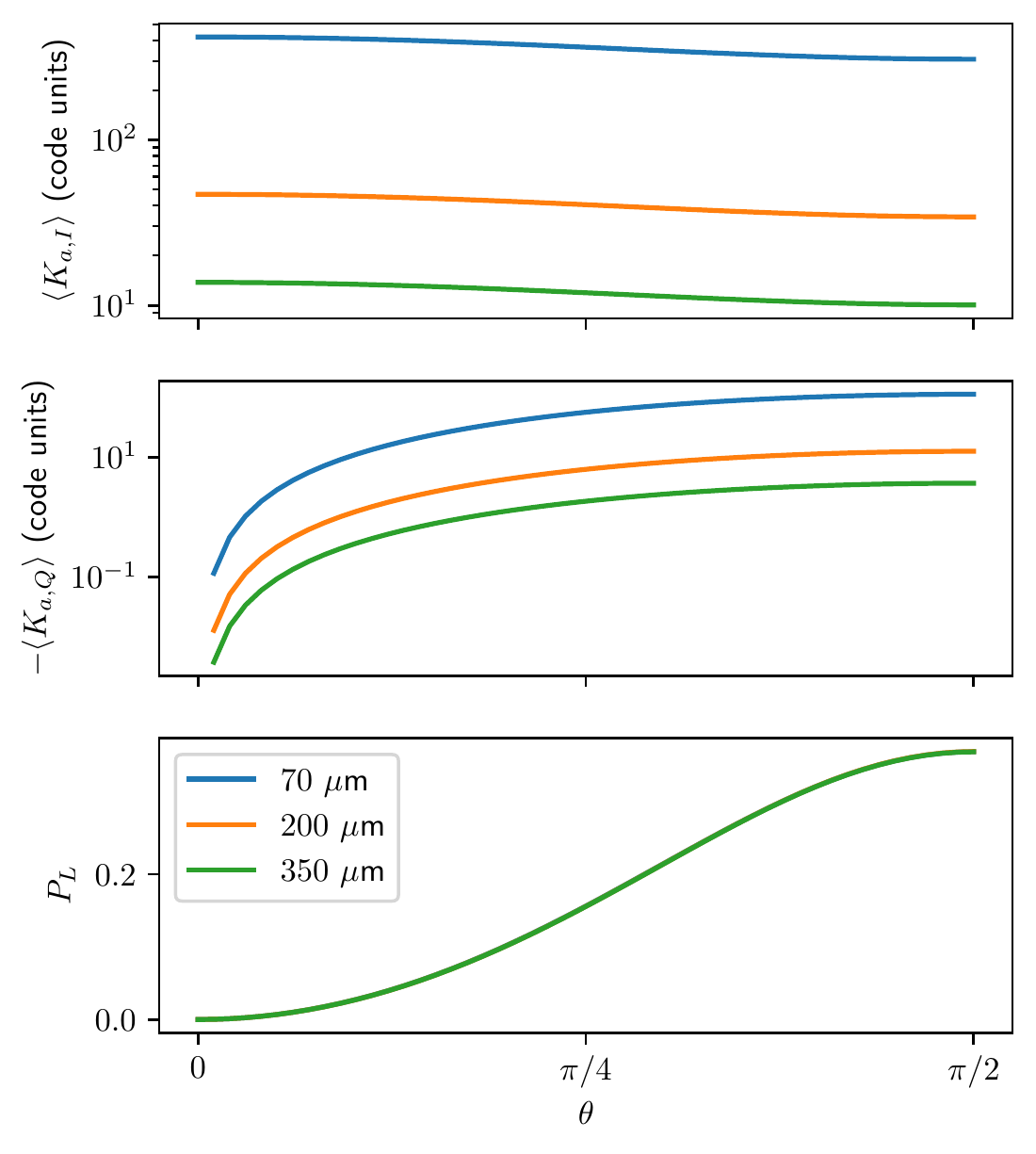}
\caption{Size-averaged absorption cross section (\emph{top}), polarized
absorption cross section (\emph{middle}) and linear polarization fraction
(\emph{bottom}) as a function of zenith angle $\theta{}$ for our fiducial
model using a CDE2 shape distribution and assuming perfect grain alignment
above a threshold grain size of $0.1~\mu{}$m. The different colors correspond
to the three different wavelengths for the planned B-BOP polarimeter.}
\label{fig:fiducial_model}
\end{figure}

The results for our fiducial model are shown in \figureref{fig:fiducial_model}.
There is no linear polarization along the direction of the magnetic field; the
linear polarization is maximal when the grains are observed in a direction
perpendicular to the magnetic field direction. The three wavelengths shown have
identical linear polarization fractions. This can be explained by the small
difference in refractive index between these three wavelengths. The total
absorption and hence emission coefficient does increase significantly at shorter
wavelengths. Below, we will hence mainly focus on results for
$\lambda{}=200~\mu{}$m.

\subsection{Alignment distribution}

Various methods to approximate the absorption coefficients for partially aligned
spheroidal grains have been discussed in literature. \citet{1974DyckBeichman}
introduced a heuristic approximation whereby a fraction $f_a$ of the grains is
perfectly aligned, and the rest has a random orientation. They do however rely
on an approximate calculation of the absorption coefficients for aligned grains,
and do not provide a recipe to relate the alignment fraction to a specific
alignment mechanism. \citet{2006Draine, 2009DraineFraisse} use so called
\emph{picket fence alignment} to compute the average extinction and absorption
cross section over all angles, whereby the alignment of the dust grains is given
in terms of a single alignment fraction $f_{\rm{}pf}$ so that a fraction
$(1+2f_{\rm{}pf})/3$ of the grains has their short axis aligned with the
magnetic field, while the remaining grains have their short axis aligned with
two mutually orthogonal directions that are both perpendicular to the magnetic
field direction in equal proportions.

This approach is similar to \citet{2016Reissl}, who go a step further and
self-consistently compute the alignment based on the local radiation field and
dust properties. They parameterize the alignment using the Rayleigh reduction
factor,
\begin{equation}
R = \left\langle{} \left(\frac{3}{2}\cos^2\beta{} - \frac{1}{2}\right)
\left(\frac{3}{2} \cos^2\zeta{} - \frac{1}{2} \right) \right
\rangle{},
\end{equation}
where $\beta{}$ is the angle between the magnetic field and the angular
momentum vector of the aligning grain, and $\zeta{}$ the angle between the
angular momentum vector and the short axis of the spheroidal grain, the so
called internal alignment angle. If we assume perfect internal alignment,
then $\zeta{}=0$ and the Rayleigh reduction factor reduces to
\begin{equation}
R = \frac{3}{2} \langle{} \cos^2\beta{} \rangle{} - \frac{1}{2},
\end{equation}
which is similar to the parametrization for alignment distributions used by
\citet{1991Mishchenko} (in fact, the Rayleigh reduction factor $R$ is equal
to twice the expansion coefficient parameter $p_2$ for Mishchenko's imperfect
alignment distribution; the factor 2 stems from a different normalization for
the angular average).

Once the Rayleigh reduction factor is known, the absorption coefficients are
given by
\begin{gather}
Q_{\rm{}abs}(\theta{}) = \langle{} Q \rangle{} + R(Q_{\parallel{}} -
Q_{\perp{}}) \left(\frac{1}{3} - \frac{1}{2}\sin^2\theta{} \right), \\
Q_{\rm{}abs,pol}(\theta{}) = -\frac{1}{2}R(Q_{\parallel{}} - Q_{\perp{}})
\sin^2\theta{},
\end{gather}
with
\begin{gather}
\langle{} Q \rangle{} = \frac{2 Q_{\perp{}} + Q_{\parallel{}}}{3}, \\
Q_{\perp{}} = Q_{\rm{}abs}\left(\theta{} = \frac{\pi{}}{2} \right), \\
Q_{\parallel{}} = Q_{\rm{}abs}(\theta{}=0),
\end{gather}
where for the last two equations we do not assume any alignment with a magnetic
field, but evaluate the absorption coefficient in a frame where the $z$ axis
corresponds to the symmetry axis of the dust grain. Note that the expression for
$\langle{}Q\rangle{}$ is different from the one given by \citet{2016Reissl} (see
also \citealp{1985LeeDraine,2009DraineFraisse}). Similarly, there is a sign
difference in the expression for $Q_{\rm{}abs,pol}$. These approximate
expressions are valid in the Rayleigh limit, i.e. for grain sizes that are
significantly smaller than the wavelength of the absorbed radiation.

If we average the expressions for $Q_{\rm{}abs}(\theta{})$ and
$Q_{\rm{}abs,pol}(\theta{})$ over all zenith angles $\theta{}$, we find
\begin{align}
\langle{} Q_{\rm{}abs} \rangle{} &= \langle{} Q \rangle{}, \\
\langle{} Q_{\rm{}abs,pol} \rangle{} &= \frac{R}{3} (Q_{\parallel{}} -
Q_{\perp{}}).
\end{align}
The Rayleigh reduction factor has the same meaning as the alignment fraction
$f_{\rm{}pf}$ for picket fence alignment \citep{2009DraineFraisse}.

We will perform two tests. In the first test, we will compute the average
absorption coefficients for a specific grain alignment using picket fence
alignment, and a full angular averaging for both the mixture of random and
perfectly aligned grains and imperfectly aligned grains. In the second test, we
will compute the zenith angle dependence of the absorption coefficients using a
mixture of random and perfect alignment, using picket fence alignment, and using
\citet{1991Mishchenko} imperfect alignment with \textsc{CosTuuM}, for a range of
alignment fractions and equivalent Rayleigh reduction factors. Since all these
methods to some extent require information about the shape (prolate or oblate)
and assume a uniform alignment mechanism for all sizes, we will show results for
a single grain size ($1~\mu{}$m) and for two fixed shapes: $d=0.5$ (prolate) and
$d=2$ (oblate).

\subsubsection{Average absorption cross section}

\begin{figure*}
\centering{}
\includegraphics[width=0.98\textwidth{}]{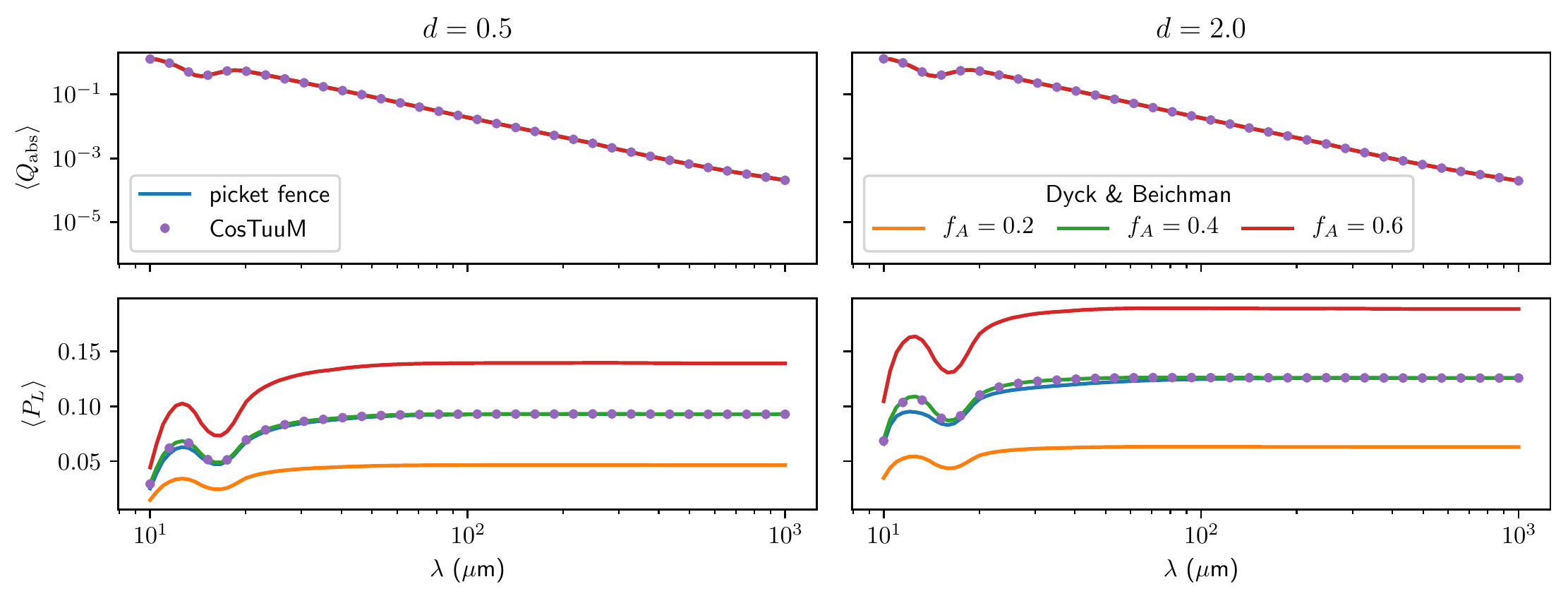}
\caption{Angular averaged absorption coefficient (\emph{top}) and linear
polarization fraction (\emph{bottom}) for prolate (\emph{left}) and oblate
(\emph{right}) grains as a function of wavelength, computed using various
methods, as indicated in the legend.}
\label{fig:alignment_averages}
\end{figure*}

\figureref{fig:alignment_averages} shows the angular averaged absorption
coefficient and linear polarization fraction for our representative grains as
a function of wavelength, calculated using the various methods. We assume all
grains are imperfectly aligned according to a Mishchenko alignment distribution
with $p_2=-0.2$ (prolate) and $p_2=0.4$ (oblate), corresponding to Rayleigh
reduction factors $R=-0.4$ and $R=0.8$ respectively. As expected, all methods
show excellent agreement for the average absorption coefficient. Both picket
fence alignment and the mixture of perfectly aligned and randomly aligned grains
provide good approximations for the average linear polarization fraction at long
wavelengths, where the Rayleigh approximation is valid. Towards shorter
wavelengths, discrepancies start to appear between the picket fence alignment
result and the \textsc{CosTuuM}-based results, as the Rayleigh approximation
starts to break down. The fully self-consistent \textsc{CosTuuM} results are
exactly reproduced by assuming a mixture with a polarized fraction $f_A=0.4$, so
a dust grain mixture with imperfectly aligned grains can be reliably reproduced
using this approximation. Note however that there is no general recipe to relate
the alignment fraction $f_A$ to the parameters of the orientation distribution,
and that computing the absorption coefficients as a function of zenith angle for
perfectly aligned grains is equally complex as computing them for an imperfectly
aligned ensemble. So this approximation has no real benefits.

\subsubsection{Zenith angle dependence}

\begin{figure*}
\centering{}
\includegraphics[width=0.98\textwidth{}]{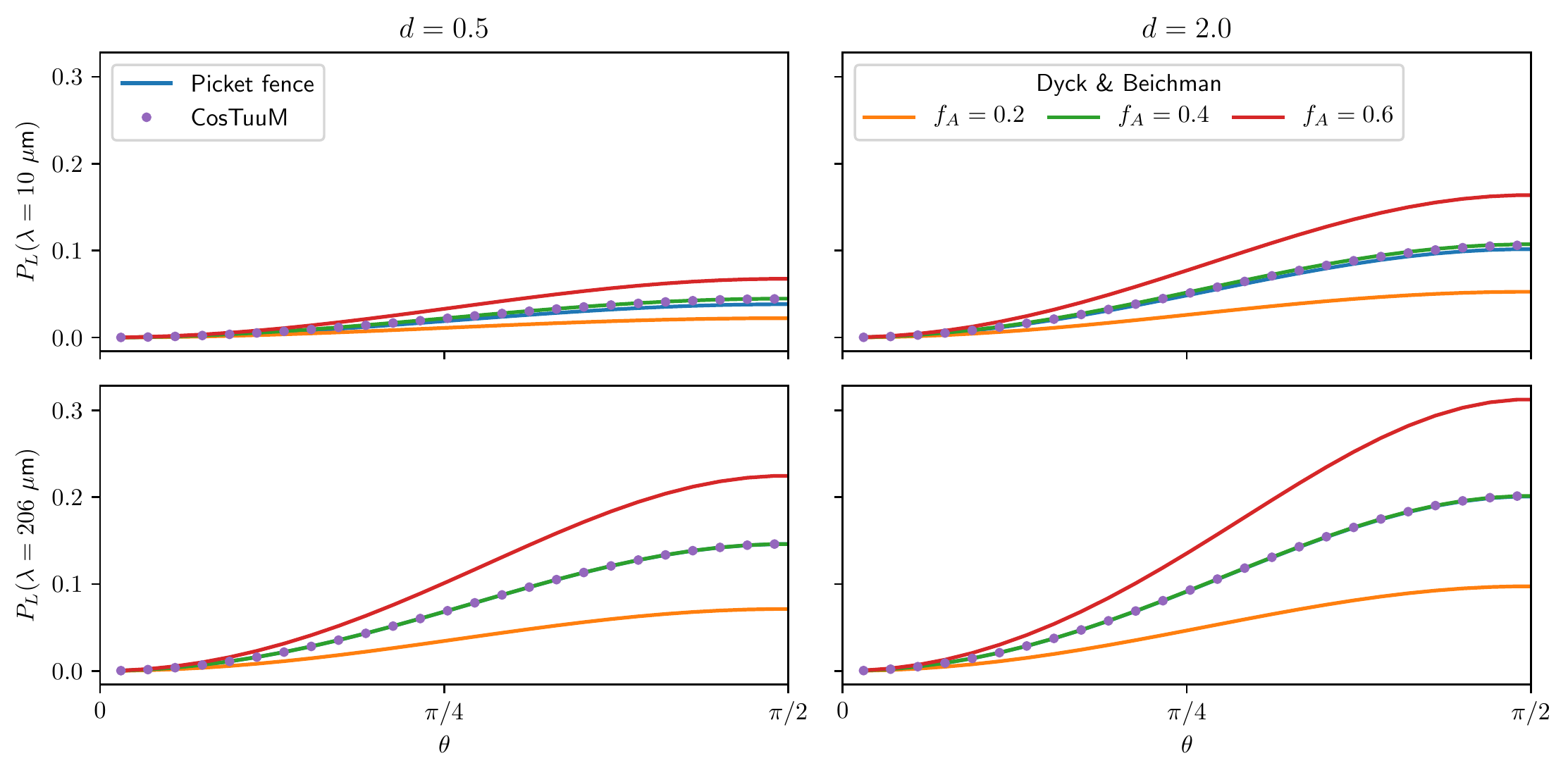}
\caption{Linear polarization fraction for prolate (\emph{left}) and oblate
(\emph{right}) grains as a function of zenith angle, computed using various
methods, as indicated in the legend. \emph{Top:} values for
$\lambda{}=10~\mu{}$m, \emph{bottom:} values for $\lambda{}=205~\mu{}$m.}
\label{fig:alignment_zenith}
\end{figure*}

\figureref{fig:alignment_zenith} shows the zenith angle dependence of the
linear polarization fraction for the same representative grains, and for two
representative wavelengths: a long wavelength for which both picket fence
alignment and a mixture of random and perfect alignment provide excellent
approximations, and a shorter wavelength for which picket fence alignment starts
to diverge. Picket fence alignment underestimates the linear polarization
fraction perpendicular to the orientation of the magnetic field, which explains
the systematic underestimation of the average linear polarization fraction we
found above.

In this case, using the \citet{1974DyckBeichman} approximation does have a clear
numerical advantage: by calculating the zenith angle dependence for the
perfectly aligned case, we can represent the correct zenith angle dependence for
any other alignment case, assuming we have a method to determine the
corresponding $f_A$.

\subsection{Shape distribution}

The shape of spheroidal dust grains is poorly constrained, and most authors
therefore tend to assume a single shape for their spheroidal dust grains, with
a preference for oblate spheroids with $d=2$
\citep{1984Draine,2006Draine,2016Reissl,2017Bertrang,2018Reissl}.
\citet{2014Siebenmorgen} explore some more extreme axis ratios
($d\in{}[2,3,4]$), while \citet{1995KimMartin} even discuss prolate grains with
$d\in{}[1/2,1/4]$. Studies that try to fit dust properties to observations tend
to prefer oblate grains with smaller axis ratios, $d=3/2$
\citep{1993DraineMalhotra}, $d=\sqrt{2}$ \citep{2006Draine}, $d\in{}[1.4,1.6]$
\citep{2009DraineFraisse}.

\citet{2003Min} and \citet{2017Draine} address the issue of shape effects
and describe the use of more general shape distributions like the CDE2
distribution we use here, but they do not explore the differences between this
full statistical sampling and the use of a single shape, nor do they show that
a single shape can in fact represent an ensemble of grains accurately.

\begin{figure}
\centering{}
\includegraphics[width=0.48\textwidth]{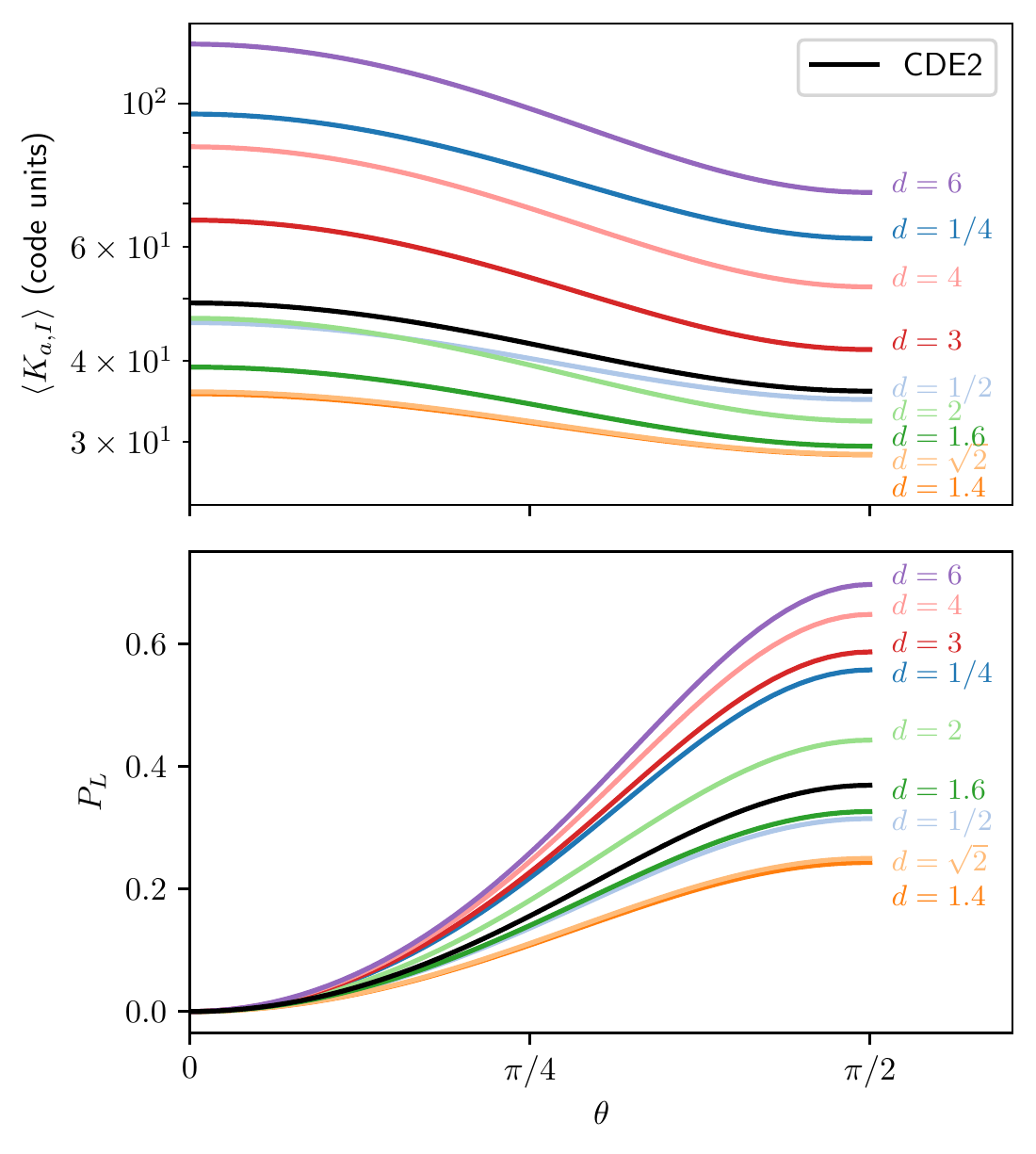}
\caption{Size-averaged absorption cross section (\emph{top}) and linear
polarization fraction (\emph{bottom}) as a function of zenith angle $\theta{}$
for various single shape grains as indicated in the figure, and for our
fiducial CDE2 shape distribution. All results are shown for a wavelength
of $200~\mu{}$m.}
\label{fig:ShapeComparison}
\end{figure}

In \figureref{fig:ShapeComparison} we show the absorption cross section and
linear polarization fraction as a function of zenith angle for our fiducial dust
grain mixture at $\lambda{}=200~\mu{}$m, and for various assumed grain shapes
and our CDE2 mixture. Perfect grain alignment is assumed. None of the assumed
single shape results matches the CDE2 mixture; while the absorption coefficient
is larger than that for a $d=2$ grain shape, the linear polarization fraction is
relatively low, in between the $d=1.6$ and $d=2$ result. Assuming a single shape
will hence underestimate the emission and overestimate the polarization. It is
also clear that the zenith angle dependence of the CDE2 distribution result is
more similar to the oblate than to the prolate grain shapes, so the assumption
of a single oblate shape is more justified. This is to be expected, as the
average shape $\langle{}d\rangle{}$ for the CDE2 distribution is
$\approx{}1.5$.

\section{Code design}
\label{sec:design}

\subsection{Program flow}
\label{sec:flow}

\begin{figure}
\centering{}
\includegraphics[width=0.49\textwidth]{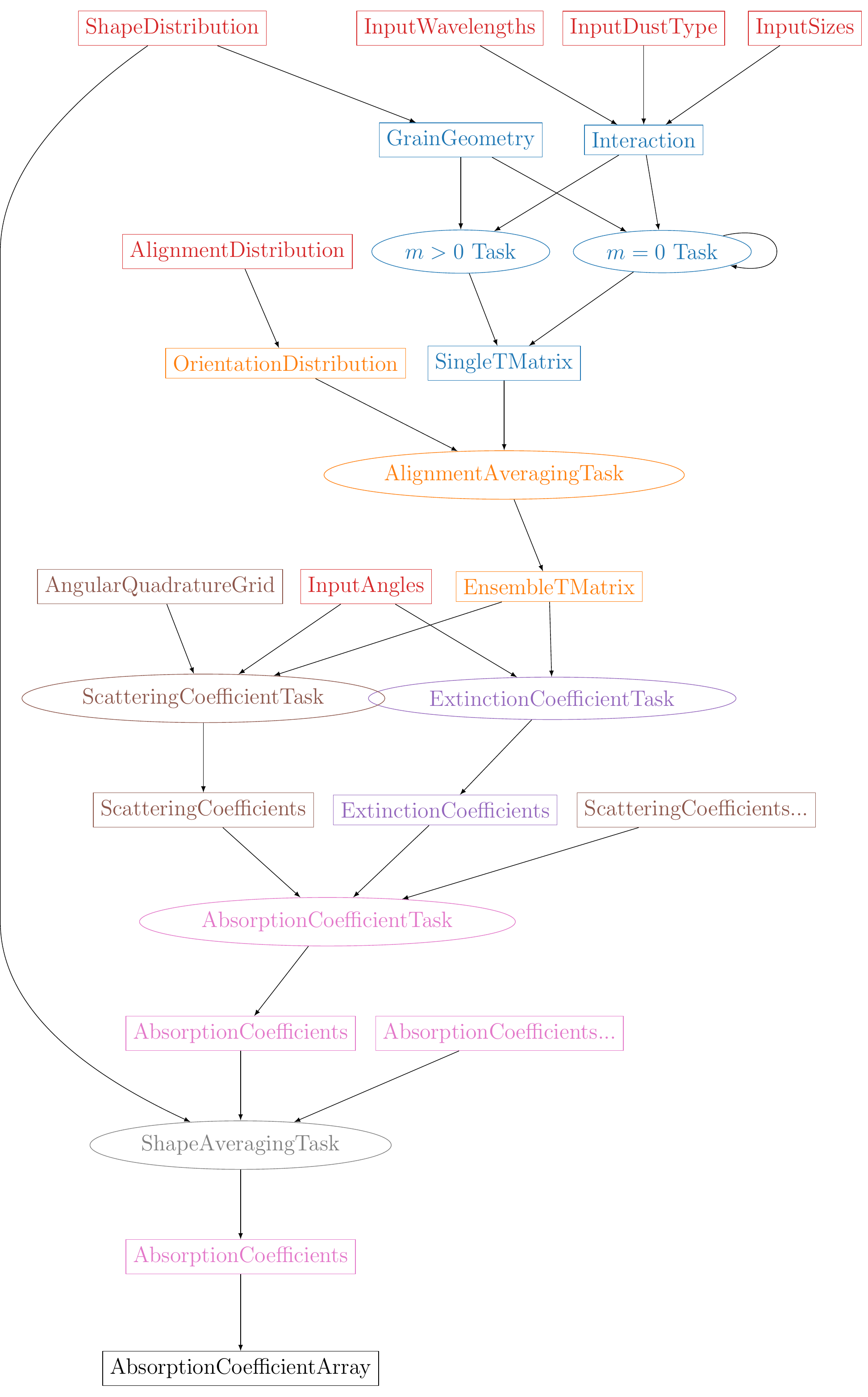}
\caption{Steps required to calculate a single set of absorption coefficients.
The output (bottom, black) is generated from six input variables (red): arrays
with grain sizes, wavelengths and output angles, a shape distribution and an
alignment distribution, and a dust grain type. The clusters of different color
correspond to different steps in the algorithm, as explained in the text.
Rectangular boxes represent data values that are stored in memory, ellipses
correspond to tasks that generate or manipulate data in memory.}
\label{fig:flowchart}
\end{figure}

\figureref{fig:flowchart} shows the dependency graph for the calculation of an
absorption coefficient table like the one presented in \sectionref{sec:results}.
This dependency graph is executed in parallel using a task-based
parallelisation algorithm. Details of this algorithm and its performance are
presented in \appendixref{sec:task_based_parallelisation}.
From bottom to top, the calculation of a single set of absorption coefficients
for a given grain size and dust type, shape distribution, alignment
distribution, incoming photon wavelength and grid of input zenith angles,
involves the following steps (the colors in between brackets match the colors in
the flowchart):

\paragraph{Shape averaging (gray)} The calculation of a single set of absorption
coefficients requires the calculation of the shape distribution average for
absorption coefficients with the same grain size and type, alignment
distribution, incoming wavelength and zenith angle grid. At the start of the
calculation, a set of representative shapes is generated. For each shape, a full
set of absorption coefficients is computed. Once all of these are available,
they are combined into a single average set of coefficients.

\paragraph{Absorption coefficient calculation (pink)} A single set of absorption
coefficients is calculated by taking the extinction coefficients for the same
input parameters and correcting for the scattering contribution for the same
parameters over a representative grid of quadrature angles according to
equation \eqref{eq:absorption}.

\paragraph{Exinction coefficients (purple)} The extinction coefficients are
calculated from equations \eqref{eq:ext_start}-\eqref{eq:ext_end} for the given
grid of input angles and using the alignment averaged ensemble T-matrix for the
given grain size, dust type, alignment distribution and incoming wavelength. We
only compute the elements of the extinction matrix that are required in the end
result.

\paragraph{Scattering coefficients (brown)} The scattering coefficients are
similarly calculated from equations \eqref{eq:sca_start}-\eqref{eq:sca_end},
again only computing the elements that we need. The numerical quadrature of the
scattering correction term requires a grid of quadrature angles; the resolution
of this grid (and hence the accuracy of the correction) is also an input
parameter (see \sectionref{sec:correction_convergence}).

\paragraph{Ensemble T-matrix (orange)} The T-matrix for a given interaction is
the result of ensemble-averaging the T-matrix for a single grain over the
orientation distribution that is determined by the given alignment distribution
and grain size. The unaveraged T-matrix still depends on the given input grain
size, grain type and incoming photon wavelength.

\paragraph{Single T-matrix (blue)} The calculation of an individual T-matrix
requires an iterative procedure whereby the order of the spherical basis
function expansion is iteratively increased until the resulting T-matrix is
sufficiently converged. To speed up this process, the procedure is performed
using a reduced version of the T-matrix ($m=0$ tasks). Once convergence is
reached, the missing components of the T-matrix can be computed ($m>0$ tasks);
the latter computation can be done in parallel for different values of the
expansion order $m$. The T-matrix calculation depends on two sets of input
parameters: the dust grain geometry which only depends on the shape (and hence
the input shape distribution), and the interaction parameters -- the input
wavelength, grain size and dust type, which together also determine
the refractive index.

\subsection{Python module}

\textsc{CosTuuM} itself has a relatively small set of functionalities, but its
results could conceivable be used in many different ways. This makes it an ideal
target to be turned into a Python library. In our design, the entire
calculation outlined in \sectionref{sec:flow} is contained within a single
library function that reads input parameters from NumPy arrays. The result of
the calculation is also returned in NumPy array format. This makes it possible
to incorporate the \textsc{CosTuuM} calculation within a script that directly
manipulates the output in any conceivable manner.

We aimed to make \textsc{CosTuuM} both easy to use for non expert users and
highly flexible for more experienced users. For this reason, most program
components (e.g. the shape distribution, the material properties...) are exposed
as separate Python objects, with specific customizable implementations that use
user-provided functions (e.g. a user-provided function that returns the
refractive index for a dust grain with given properties). These classes are
hidden by sensible defaults from users that are not as familiar with the way the
library works.

All of the figures in this work (except for
\figureref{fig:reference_frame}) where generated using the \textsc{CosTuuM}
Python library and are included within the public repository for the code.
Potential users are encouraged to use these scripts as a reference when creating
their own \textsc{CosTuuM} based scripts.

\section{Conclusion}
\label{sec:conclusion}

In this work, we presented the open-source software package \textsc{CosTuuM}
that can be used to compute absorption and emission coefficients for arbitrary
mixtures of spheroidal dust grains that are (partially) aligned with a magnetic
field. We showed that \textsc{CosTuuM} reproduces widely used results from
literature, and that is fit for its purpose: providing accurate optical
properties for use in radiative transfer applications. We emphasized this point
by predicting the expected polarization signal for perfectly aligned silicate
grains at the wavelengths for the B-BOP polarimeter aboard the proposed ESA/JAXA
mission SPICA.

We used \textsc{CosTuuM} to test the validity of some common approximations.
We first looked at the absorption cross sections and linear
polarization fraction for an ensemble of dust grains that all have the same
shape and size, and that are aligned with a magnetic field according to a known
alignment distribution. We find that the absorption coefficients and linear
polarization function computed using \textsc{CosTuuM} can be accurately
reproduced using a mixture of randomly oriented and perfectly aligned grains,
and that both techniques are more accurate at short wavelengths than the widely
used picket fence alignment approximation. This finding is important
for applications where grain alignment depends on spatial position, since it
makes it possible to obtain all properties using a combination of just two
pre-computed tables, provided that an appropriate polarized fraction can be
defined.

Next, we compared the absorption cross section and linear
polarization fraction as a function of zenith angle for different grain shape
distributions, and assuming an ensemble of grains with a realistic size
distribution. We find that a realistic grain shape distribution consisting of
both oblate and prolate grains has a similar zenith angle dependence for both
quantities as a representative oblate grain with a fixed shape. We however also
find that it is not possible to represent both quantities simultaneously with a
single representative grain. While the ensemble average absorption coefficient
for the mixture is similar to that of oblate grains with relatively large axis
ratios $d>2$, the matching linear polarization fraction is more similar to
oblate grains with moderate axis ratios ($d\approx{}1.6$). The discrepancies we
find between various single shape approximations are much more significant than
the differences between self-consistent alignment and picket fence alignment.
Insufficient knowledge of an appropriate shape distribution is hence the main
obstacle in obtaining accurate optical properties.

\section*{Acknowledgments}

The authors acknowledge financial support from the Belgian Science 
Policy Office (BELSPO) through the PRODEX project ``SPICA-SKIRT: A 
far-infrared photometry and polarimetry simulation toolbox in 
preparation of the SPICA mission'' (C4000128500).

\software{Matplotlib \citep{Matplotlib}, NumPy \citep{NumPy}, SciPy
\citep{SciPy}}

\appendix{}
\restartappendixnumbering{}

\section{T-matrix calculation}

\subsection{The T-matrix}
\label{appendix:Tmatrix}

The T-matrix method was first introduced by \citet{1971Waterman} and
consequently adopted by \citet{1984Tsang, 1998Mishchenko} and others, see
\citet{2018Doicu} for a recent derivation. Below, we give a practical summary
of the method as we have implemented it in \textsc{CosTuuM}.

The calculation of the T-matrix for a single scatterer is based on an expansion
of the incoming, local and scattered electromagnetic waves in spherical basis
functions. These basis functions are the general solutions of the Helmholtz
equation
\begin{equation}
\nabla{}^2 \boldsymbol{A} + k^2 \boldsymbol{A} = 0
\end{equation}
in spherical coordinates, and are given by
\begin{align}
\boldsymbol{M}_{mn}(r, \theta{}, \phi{}) &= \nabla{} \times{}
\boldsymbol{A}_{mn}(r, \theta{}, \phi{}), \\
\boldsymbol{N}_{mn}(r, \theta{}, \phi{}) &= \frac{1}{k} \nabla{} \times{} \left(
\nabla{} \times{} \boldsymbol{A}_{mn}(r, \theta{}, \phi{}) \right), \\
{\rm{}Rg}\boldsymbol{M}_{mn}(r, \theta{}, \phi{}) &= \nabla{} \times{} {\rm{}Rg}
\boldsymbol{A}_{mn}(r, \theta{}, \phi{}), \\
{\rm{}Rg}\boldsymbol{N}_{mn}(r, \theta{}, \phi{}) &= \frac{1}{k} \nabla{}
\times{} \left( \nabla{} \times{} {\rm{}Rg} \boldsymbol{A}_{mn}(r, \theta{},
\phi{}) \right),
\end{align}
with
\begin{align}
\boldsymbol{A}_{mn}(r, \theta{}, \phi{}) &= B_{mn}(\theta{},\phi{})
h^{(1)}_n(kr) \boldsymbol{r}, \\
{\rm{}Rg} \boldsymbol{A}_{mn}(r, \theta{}, \phi{}) &= B_{mn}(\theta{},\phi{})
j^{(1)}_n(kr) \boldsymbol{r}
\end{align}
and
\begin{equation}
B_{mn} (\theta{}, \phi{}) = (-1)^m \frac{1}{\sqrt{4\pi{}}}
\sqrt{\frac{2n+1}{n(n+1)}} {\rm{}e}^{im\phi{}} d^n_{0m}(\theta{}).
\end{equation}
In these expressions, the radial dependence is given by the spherical Hankel
and Bessel functions of the first kind, $h^{(1)}_n(x)$ and $j^{(1)}_n(x)$,
while the dependence on the zenith angle $\theta{}$ is given by the Wigner
d-functions
\begin{equation}
d^n_{0m}(\theta{}) = (-1)^{m} \sqrt{\frac{(n-m)!}{(n+m)!}}
P^m_n(\cos\theta{}),
\end{equation}
with $P^m_n(x)$ the associated Legendre polynomial of degree $n$ and order $m$.
The expansion in theory goes up to $n=\infty{}$, but in practice can usually be
truncated at some finite $n_{\rm{}max}$. The order $m$ in these expressions has
values $m \in{} [-n, n]$.

Note that the basis functions $[\boldsymbol{M},\boldsymbol{N}]_{mn}$ and their
regular (${\rm{}Rg}$) counterparts satisfy the following relations:
\begin{align}
\frac{1}{k} \nabla{} \times{} \boldsymbol{N}_{mn} &= \boldsymbol{M}_{mn},\\
\frac{1}{k} \nabla{} \times{} \boldsymbol{M}_{mn} &= \boldsymbol{N}_{mn},
\end{align}
which can be easily verified using the generating Helmholtz equation.

To calculate the T-matrix, the incoming (plane) electromagnetic wave is expanded
in terms of the spherical basis functions
${\rm{}Rg}[\boldsymbol{M},\boldsymbol{N}]_{mn}(kr,\theta{},\phi{})$, while the
scattered (spherical) wave is expanded in terms of the
$[\boldsymbol{M},\boldsymbol{N}]_{mn}(kr,\theta{},\phi{})$. The T-matrix or
\emph{transition matrix} is the matrix that links the expansion coefficients in
both expansions. Following \citet{1971Waterman}, we can obtain the T-matrix
by requiring that the tangential components of the electromagnetic field are
continuous across the \emph{boundaries} between the vacuum and the scattering
particle. To this end, we also need the expansion of the \emph{internal}
electromagnetic field in terms of
${\rm{}Rg}[\boldsymbol{M},\boldsymbol{N}]_{mn}(k_rr,\theta{},\phi{})$, where
$k_r = km_r$ is now a generalized wavenumber that also includes the (complex)
refractive index of the scattering particle, $m_r$.

Within this so called \emph{extended boundary condition} framework, the T-matrix
is given by
\begin{equation}
\mathbf{T} = -{\rm{}Rg}\mathbf{Q} \mathbf{Q}^{-1},
\end{equation}
where the matrix $Q$ is given by
\begin{equation}
\mathbf{Q} = \begin{pmatrix}
\mathbf{Q}^{MM} & \mathbf{Q}^{MN} \\
\mathbf{Q}^{NM} & \mathbf{Q}^{NN}
\end{pmatrix},
\end{equation}
with
\begin{multline}
Q^{XY}_{mnm'n'} = k \int{} {\rm{}d}\boldsymbol{\sigma{}} . \left[ \right.\\ \left.
\left( \nabla{} \times{} {\rm{}Rg}\boldsymbol{X}_{m'n'}(k_rr, \theta{}, \phi{})
\right) \times{} \boldsymbol{Y}_{mn}(kr, \theta{}, \phi{}) \right. \\ \left. +
{\rm{}Rg}\boldsymbol{X}_{m'n'}(k_rr, \theta{}, \phi{}) \times{} \left( \nabla{}
\times{} \boldsymbol{Y}_{mn}(kr, \theta{}, \phi{}) \right) \right],
\label{equation:Q}
\end{multline}
with the surface element in this expression given by
\begin{equation}
{\rm{}d}\boldsymbol{\sigma{}} = r^2 \sin\theta{} {\rm{}d}\theta{}
{\rm{}d}\phi{} \left( \hat{r} -
\frac{1}{r}\frac{{\rm{}d}r(\theta{})}{{\rm{}d}\theta{}} \hat{\theta{}} \right),
\end{equation}
where $\hat{r}$ and $\hat{\theta{}}$ represent the radial and zenithal unit
vectors in the spherical coordinate system, and $r(\theta{})$ is the equation
that describes the surface of the spheroid. To get the regular matrix
${\rm{}Rg}\mathbf{Q}$, it suffices to replace the non-regular basis functions
with their regular counterparts in these expressions.

The full expressions for the matrix elements $Q^{XY}$ can be derived by
substituting the expressions for the basis functions in equation
\eqref{equation:Q} and carrying out the trivial integration over the azimuthal
angle $\phi{}$:
\begin{equation}
\int_0^{2\pi{}} {\rm{}e}^{i(m-m')\phi{}} {\rm{}d}\phi{} = 2\pi{}\delta{}_{mm'},
\end{equation}
with $\delta{}_{mm'}$ the Kronecker delta. We can precompute \citep[based
on][]{1984Tsang}
\begin{multline}
(MM)_{m_1n_1m_2n_2} = \\
\begin{gathered}
\int{} {\rm{}d}\boldsymbol{\sigma{}} . \left[
\boldsymbol{M}_{m_1n_1}(k_1r,\theta{},\phi{}) \times{}
\boldsymbol{M}_{m_2n_2}(k_2r,\theta{},\phi{}) \right] = \\
-\frac{i}{2} C_{n_1n_2} (-1)^{m_1+m_2} \delta{}_{m_1-m_2} \\
\int_0^{\pi{}} r^2
\sin\theta{} h^{(1)}_{n_1}(k_1r) h^{(1)}_{n_2}(k_2r) \\
\left[\pi{}_{m_1n_1}(\theta{}) \tau{}_{-m_1n_2}(\theta{}) \right. \\ \left. -
\tau{}_{m_1n_1}(\theta{}) \pi{}_{-m_1n_2}(\theta{}) \right] {\rm{}d}\theta{},
\end{gathered}
\end{multline}
\begin{multline}
(MN)_{m_1n_1m_2n_2} = \\
\begin{gathered}
\int{} {\rm{}d}\boldsymbol{\sigma{}} . \left[
\boldsymbol{M}_{m_1n_1}(k_1r,\theta{},\phi{}) \times{}
\boldsymbol{N}_{m_2n_2}(k_2r,\theta{},\phi{}) \right] = \\
\frac{1}{2} C_{n_1n_2} (-1)^{m_1+m_2} \delta{}_{m_1-m_2} \\
\int_0^{\pi{}} r^2
\sin\theta{} \Bigg( h^{(1)}_{n_1}(k_1r) {\rm{}D}h^{(1)}_{n_2}(k_2r) \\
\left[-\pi{}_{m_1n_1}(\theta{}) \pi{}_{-m_1n_2}(\theta{}) +
\tau{}_{m_1n_1}(\theta{}) \tau{}_{-m_1n_2}(\theta{}) \right] \\ +
\frac{1}{r}\frac{{\rm{}d}r(\theta{})}{{\rm{}d}\theta{}}
\frac{n_2(n_2 + 1)}{k_2r} \\
h^{(1)}_{n_1}(k_1r) h^{(1)}_{n_2}(k_2r)
\tau{}_{m_1n_1}(\theta{}) d^{n_2}_{0-m_1}(\theta{})
\Bigg)
{\rm{}d}\theta{}\\
= -\int{} {\rm{}d}\boldsymbol{\sigma{}} . \left[
\boldsymbol{N}_{m_2n_2}(k_2r,\theta{},\phi{}) \times{}
\boldsymbol{M}_{m_1n_1}(k_1r,\theta{},\phi{}) \right]
\end{gathered}
\\= -(NM)_{m_2n_2m_1n_1},
\end{multline}
\begin{multline}
(NN)_{m_1n_1m_2n_2} = \\
\begin{gathered}
\int{} {\rm{}d}\boldsymbol{\sigma{}} . \left[
\boldsymbol{N}_{m_1n_1}(k_1r,\theta{},\phi{}) \times{}
\boldsymbol{N}_{m_2n_2}(k_2r,\theta{},\phi{}) \right] = \\
-\frac{i}{2} C_{n_1n_2} (-1)^{m_1+m_2} \delta{}_{m_1-m_2} \\
\int_0^{\pi{}} r^2
\sin\theta{} \Big( {\rm{}D}h^{(1)}_{n_1}(k_1r) {\rm{}D}h^{(1)}_{n_2}(k_2r) \\
\left[ -\tau{}_{m_1n_1}(\theta{}) \pi{}_{-m_1n_2}(\theta{}) +
\pi{}_{m_1n_1}(\theta{}) \tau{}_{-m_1n_2}(\theta{}) \right]\\ +
\frac{1}{r}\frac{{\rm{}d}r(\theta{})}{{\rm{}d}\theta{}} \Big[
\frac{n_2(n_2 + 1)}{k_2r} {\rm{}D}h^{(1)}_{n_1}(k_1r) h^{(1)}_{n_2}(k_2r) \\
\pi{}_{m_1n_1}(\theta{}) d^{n_2}_{0-m_1}(\theta{}) - \\
\frac{n_1(n_1 + 1)}{k_1r} h^{(1)}_{n_1}(k_1r) {\rm{}D}h^{(1)}_{n_2}(k_2r) \\
d^{n_1}_{0m_1}(\theta{}) \pi{}_{-m_1n_2}(\theta{})
\Big] \Bigg) {\rm{}d}\theta{},
\end{gathered}
\end{multline}
with
\begin{align}
C_{nn'} &= \sqrt{\frac{(2n+1)(2n'+1)}{n(n+1)n'(n'+1)}}, \\
\pi{}_{mn}(\theta{}) &= \frac{m}{\sin(\theta{})} d^n_{0m}(\theta{}),
\label{eq:pi} \\
\tau{}_{mn}(\theta{}) &= \frac{d}{d\theta{}} d^n_{0m}(\theta{}),
\label{eq:tau} \\
{\rm{}D}h^{(1)}_{n}(x) &= \frac{[xh^{(1)}_n(x)]'}{x} = \frac{{\rm{}d}}{{\rm{}d}x}
h^{(1)}_n(x) - \frac{1}{x} h^{(1)}_n(x).
\end{align}
In these expressions, the spherical Hankel functions $h^{(1)}_n(x)$ are to be
replaced with the corresponding spherical Bessel functions $j^{(1)}_n(x)$ where
appropriate when substituting ${\rm{}Rg}[M,N]$ for $[M,N]$.

To carry out the second integration over the zenith angle $\theta{}$, we follow
\citet{1984Tsang, 1998Mishchenko} and use a numerical Gauss-Legendre quadrature
in the variable $\cos(\theta{})$. Note that to derive the expressions above, we
have made use of the relation
\begin{equation}
d^n_{0m}(\theta{}) = (-1)^m d^n_{0-m}(\theta{})
\end{equation}
and its derivative.

The final expressions for the elements of the matrix $\mathbf{Q}$ are then
\begin{align}
Q^{MM}_{mnm'n'} &= kk_r ({\rm{}Rg}NM)_{mnm'n'} + k^2 ({\rm{}Rg}MN)_{mnm'n'}, \\
Q^{MN}_{mnm'n'} &= kk_r ({\rm{}Rg}MM)_{mnm'n'} + k^2 ({\rm{}Rg}NN)_{mnm'n'}, \\
Q^{NM}_{mnm'n'} &= kk_r ({\rm{}Rg}NN)_{mnm'n'} + k^2 ({\rm{}Rg}MM)_{mnm'n'}, \\
Q^{NN}_{mnm'n'} &= kk_r ({\rm{}Rg}MN)_{mnm'n'} + k^2 ({\rm{}Rg}NM)_{mnm'n'}.
\end{align}

The T-matrix for a spheroid in a reference frame where the vertical axis is
aligned with the rotation axis of the spheroid has a number of properties,
including
\begin{align}
T^{(ij)}_{mnm'n'} &= \delta{}_{mm'} T^{(ij)}_{mnmn'} \\
T^{(ij)}_{mnmn'} &= T^{(ji)}_{-mn-mn'} = (-1)^{i+j} T^{(ij)}_{-mn-mn'},
\label{eq:Tmatrix_symmetry}
\end{align}
where $i=1,2$ and $j=1,2$ represent the $\boldsymbol{M}$ and $\boldsymbol{N}$
blocks in the T-matrix respectively. The former relation means that the T-matrix
is a block diagonal matrix in the index $m$, which allows us to compute it
separately for different values of $m$, while the latter relation means we only
need to compute the matrix elements for positive values of $m$. As in
\citet{1998Mishchenko}, this leads to the following general approach to
computing the T-matrix for a single spheroidal particle:

\begin{enumerate}

\item{} Compute the $m=0$ component of the T-matrix for a suitable guess of
$n_{\rm{}max}$ and the number of Gauss-Legendre quadrature points $n_{\rm{}GL}$.
Compute the average extinction and scattering cross sections for randomly
oriented grains \citep{1998Mishchenko}
\begin{align}
\langle{} C_{\rm{}ext} \rangle{} &= -\frac{2\pi{}}{k^2}
\sum_{n=1}^{n_{\rm{}max}} \sum_{m=-n}^n \sum_{i=1}^2
{\rm{}Re}\left(T^{(ii)}_{mnmn}\right),\\
\langle{} C_{\rm{}sca} \rangle{} &= \frac{2\pi{}}{k^2} \sum_{n=1}^{n_{\rm{}max}}
\sum_{n'=1}^{n_{\rm{}max}} \sum_{m=-n}^n \sum_{m'=-n'}^{n'} \sum_{i=1}^2
\sum_{j=1}^{2} \left| T^{(ij)}_{mnm'n'} \right|^2
\end{align}
for this sparse version of the T-matrix, where ${\rm{}Re}(z)$ represents the
real part of the complex number $z$, while $|z|^2 = zz^*$ is its norm. Repeat
this for $n'_{\rm{}max} = n_{\rm{}max}+1$ until the relative difference between
the respective values for $\langle{} C_{\rm{}ext} \rangle{}$ and $\langle{}
C_{\rm{}sca} \rangle{}$ for two successive $n_{\rm{}max}$ values drops below
some desired tolerance, $\tau{}$ (we use $\tau{} = 10^{-4}$). This determines
$n_{\rm{}max}$.

\item{} Now increase the number of Gauss-Legendre quadrature points
$n'_{\rm{}GL} = n_{\rm{}GL} + 1$ and repeat this until the relative differences
for $\langle{} C_{\rm{}ext} \rangle{}$ and $\langle{} C_{\rm{}sca} \rangle{}$
again drop below $\tau{}$. This determines $n_{\rm{}GL}$. Note that we
generally do not perform this step in \textsc{CosTuuM} because it makes it harder to
efficiently parallelize the algorithm. Instead, we assume a fixed ratio
$n_{\rm{}max} / n_{\rm{}GL} = 2$, which converges for all our tests.

\item{} Now use these values of $n_{\rm{}max}$ and $n_{\rm{}GL}$ to compute
the additional T-matrix elements for $m>0$. The elements for $m<0$ can be found
using equation \eqref{eq:Tmatrix_symmetry} and are not used in further
calculations for that same reason.

\end{enumerate}

\subsection{Orientational averages}
\label{sec:orientational_averages}

As shown in \citet{1991Mishchenko}, it is possible to compute the T-matrix
for an ensemble of spheroidal grains with different orientations w.r.t. a
reference axis. Given the orientation distribution for the ensemble of grains
(normalized according to equation \eqref{eq:orientation_distribution_norm}), tke
T-matrix for the ensemble can be computed as
\begin{multline}
\langle{}T^{(ij)}_{mnmn'} \rangle{} =
\sum_{m_1=-{\rm{}min}(n,n')}^{{\rm{}min}(n,n')} \sum_{n_1=|n-n'|}^{n+n'}\\
(-1)^{m+m_1} p_{n_1} C^{n_10}_{nmn'-m} C^{n_10}_{nm_1n'-m_1}
T^{(ij)}_{m_1nm_1n'},
\end{multline}
where $C^{NM}_{n_1m_1n_2m_2}$ are the Clebsch-Gordan coefficients and
\begin{equation}
p_n = \int_0^{\pi{}} p(\beta{}) d^n_{00}(\beta{}) \sin\beta{} {\rm{}d}\beta{}
\end{equation}
are the coefficients of the expansion of the orientation distribution in
Legendre polynomials.

Due to the orthogonality of the Legendre polynomials, the normalization
condition \eqref{eq:orientation_distribution_norm} can be guaranteed by setting
$p_0=1$. The Mishchenko distribution used in
\sectionref{sec:Mishchenko_benchmark} can be recovered by setting all
coefficients other than $p_0$ and $p_2$ to 0.

\subsection{Parity rules}

The calculation of the T-matrix elements can be significantly simplified by
taking into account the following parity rules:
\begin{align}
d^n_{0m}(\pi{}-\theta{}) &= (-1)^{n+m} d^n_{0m}(\theta{}), \\
\pi{}_{mn}(\pi{}-\theta{}) &= (-1)^{n+m} d^n_{0m}(\theta{}), \\
\tau_{mn}(\pi{}-\theta{}) &= (-1)^{n+m+1} \tau{}_{mn}(\theta{}),
\end{align}
which can all be derived from the parity rule for the associated Legendre
polynomials:
\begin{equation}
P^m_n(-x) = (-1)^{n+m} P^m_n(x).
\end{equation}

Over the integration interval $[0,\pi{}]$, this means that the
$(MN)_{m_1n_1m_2n_2}$ elements are only non-zero for even values of $n_1+n_2$,
while the $(MM)_{m_1n_1m_2n_2}$ and $(NN)_{m_1n_1m_2n_2}$ elements are only
non-zero for odd values. Both of these results are irrespective of the values of
$m_1$ and $m_2$ because of the appearance of the Kronecker delta in the
expressions for the matrix elements.

Similar rules can be derived w.r.t. the sign of $m$:
\begin{align}
\pi{}_{-mn}(\theta{}) &= (-1)^{m+1} \pi{}_{mn}(\theta{}), \\
\tau{}_{-mn}(\theta{}) &= (-1)^m \tau{}_{mn}(\theta{}).
\end{align}
These will prove useful below.

\section{Scattering events}

\subsection{The forward scattering matrix}

In a reference frame where the rotation axis of the spheroidal particle is
aligned with the vertical axis, the $\mathbf{S}$ matrix can be computed using
\citep{2000Mishchenko}
\begin{equation}
\begin{gathered}
S_{\theta{}\theta{}} =
\frac{1}{k} \sum_{n=1}^{n_{max}} \sum_{n'=1}^{n_{max}} \sum_{m = -\min(n,
n')}^{\min(n, n')} C'_{nn'} {\rm{}e}^{im(\phi{}_s^P - \phi{}_i^P)} \Big[ \\
T^{11}_{mnmn'} \pi{}_{mn}(\theta{}_s^P) \pi{}_{mn'}(\theta{}_i^P) \\
+ T^{21}_{mnmn'} \tau{}_{mn}(\theta{}_s^P) \pi{}_{mn'}(\theta{}_i^P) \\
+ T^{12}_{mnmn'} \pi{}_{mn}(\theta{}_s^P) \tau{}_{mn'}(\theta{}_i^P) \\ 
+ T^{22}_{mnmn'} \tau{}_{mn}(\theta{}_s^P) \tau{}_{mn'}(\theta{}_i^P) \Big],
\end{gathered}
\end{equation}
\begin{equation}
\begin{gathered}
S_{\theta{}\phi{}} =
-\frac{i}{k} \sum_{n=1}^{n_{max}} \sum_{n'=1}^{n_{max}} \sum_{m = -\min(n,
n')}^{\min(n, n')} C'_{nn'} {\rm{}e}^{im(\phi{}_s^P - \phi{}_i^P)} \Big[ \\
T^{11}_{mnmn'} \pi{}_{mn}(\theta{}_s^P) \tau{}_{mn'}(\theta{}_i^P) \\
+ T^{21}_{mnmn'} \tau{}_{mn}(\theta{}_s^P) \tau{}_{mn'}(\theta{}_i^P) \\
+ T^{12}_{mnmn'} \pi{}_{mn}(\theta{}_s^P) \pi{}_{mn'}(\theta{}_i^P) \\ 
+ T^{22}_{mnmn'} \tau{}_{mn}(\theta{}_s^P) \pi{}_{mn'}(\theta{}_i^P) \Big],
\end{gathered}
\end{equation}
\begin{equation}
\begin{gathered}
S_{\phi{}\theta{}} = 
\frac{i}{k} \sum_{n=1}^{n_{max}} \sum_{n'=1}^{n_{max}} \sum_{m = -\min(n,
n')}^{\min(n, n')} C'_{nn'} {\rm{}e}^{im(\phi{}_s^P - \phi{}_i^P)} \Big[ \\
T^{11}_{mnmn'} \tau{}_{mn}(\theta{}_s^P) \pi{}_{mn'}(\theta{}_i^P) \\
+ T^{21}_{mnmn'} \pi{}_{mn}(\theta{}_s^P) \pi{}_{mn'}(\theta{}_i^P) \\
+ T^{12}_{mnmn'} \tau{}_{mn}(\theta{}_s^P) \tau{}_{mn'}(\theta{}_i^P) \\ 
+ T^{22}_{mnmn'} \pi{}_{mn}(\theta{}_s^P) \tau{}_{mn'}(\theta{}_i^P) \Big],
\end{gathered}
\end{equation}
\begin{equation}
\begin{gathered}
S_{\phi{}\phi{}} =
\frac{1}{k} \sum_{n=1}^{n_{max}} \sum_{n'=1}^{n_{max}} \sum_{m = -\min(n,
n')}^{\min(n, n')} C'_{nn'} {\rm{}e}^{im(\phi{}_s^P - \phi{}_i^P)} \Big[ \\
T^{11}_{mnmn'} \tau{}_{mn}(\theta{}_s^P) \tau{}_{mn'}(\theta{}_i^P) \\
+ T^{21}_{mnmn'} \pi{}_{mn}(\theta{}_s^P) \tau{}_{mn'}(\theta{}_i^P) \\
+ T^{12}_{mnmn'} \tau{}_{mn}(\theta{}_s^P) \pi{}_{mn'}(\theta{}_i^P) \\ 
+ T^{22}_{mnmn'} \pi{}_{mn}(\theta{}_s^P) \pi{}_{mn'}(\theta{}_i^P) \Big],
\end{gathered}
\end{equation}
with
\begin{equation}
C'_{nn'} = i^{n' - n -1} C_{nn'} = i^{n' - n -1}
\sqrt{\frac{(2n+1)(2n'+1)}{n(n+1)n'(n'+1)}},
\end{equation}
and $\pi{}_{mn}(\theta{})$ and $\tau{}_{mn}(\theta{})$ as given in equations
\eqref{eq:pi} and \eqref{eq:tau}.

Using the symmetry relations for the T-matrix and the rules given above, it can
be shown that the expressions in between brackets for $S_{\theta{}\theta{}}$ and
$S_{\phi{}\phi{}}$ are even under a $m\rightarrow{}-m$ substitution, while the
expressions in between brackets for $S_{\theta{}\phi{}}$ and
$S_{\phi{}\theta{}}$ are odd. Combined with the respective odd and even nature
of the real and imaginary parts of $e^{im(\phi{}_s^P - \phi{}_i^P)} =
\cos\left(m(\phi{}_s^P - \phi{}_i^P)\right) + i \sin\left(m(\phi{}_s^P -
\phi{}_i^P)\right)$, this allows us to significantly simplify the calculation
of the forward scattering matrix elements. By rearranging the equations further
so that the sum over $m$ becomes the outer loop, we can additionally compute the
exponential factor recursively, so that only two (expensive) evaluations of
trigonometric functions are required per scattering interaction.

\subsection{Sampling grid}
\label{sec:sampling_grid}

When \textsc{CosTuuM} is used to generate tables containing absorption coefficients,
these coefficients are computed on a regular angular grid in $\theta{}_i$.
Because of the azimuthal symmetry of spheroidal grains, an arbitrary input
zenith angle can be used, we choose $\phi{}_i=0$. To correct for the scattering
contribution, we need to subtract the relevant scattering matrix components,
integrated over all incoming scattering angles. We will use a 2D grid of
Gauss-Legendre quadrature points in $\phi{}, \cos\theta{}$ for this.

In the end, we are hence left with a small grid of angles that are known at the
start of the computation. As an optimization, we can precompute the
trigonometric functions and related functions (i.e. the Wigner d-functions and
derivatives) on this grid, and use grid indices rather than angles to obtain the
relevant function values when needed. Surprisingly, this approach does not
result in a significant speed gain on our test systems, indicating that memory
bandwidth issues rather than processor speed are determining the overall cost
of the algorithm.

\section{Task-based parallelisation}
\label{sec:task_based_parallelisation}

\subsection{Algorithm}

The overall structure of the T-matrix algorithm (as illustrated in
\figureref{fig:flowchart}) makes this algorithm perfectly suited for a
task-based parallelisation strategy \citep{2016Gonnet}, a parallel programming
paradigm that has gained recent interest in the astrophysical simulation
community \citep{2012Bordner, 2016Schaller, 2016White, 2018Nordlund}. Within
this paradigm, the entire calculation is broken down into small chunks
(\emph{tasks}) that can be executed concurrently. These tasks act on
\emph{resources}, i.e. structured memory blocks that contain intermediate
results and are stored in a large shared memory pool. Tasks that require access
to the same resource cannot be executed simultaneously by different threads,
while tasks that depend on the result of another task need to be executed after
that other task finishes. These dependencies can be derived from the
dependency graph in \figureref{fig:flowchart}.

Some components required by the calculation outlined in
\figureref{fig:flowchart} can be reused between tasks. If for example two values
for one of the input parameters (e.g. the grain size) are given, then the shape
averaging procedure has to be done twice: once for every value of the grain
size. The quadrature points used to sample the shape distribution can in this
case be reused and the dust grain geometries, which only depend on these
quadrature points, only need to be computed once. Similarly, the quadrature
grid used to compute the scattering correction can be shared between all
scattering coefficient calculations. The calculation of a single T-matrix
contains a number of precomputed special functions that are not shown in
\figureref{fig:flowchart} and that can also be reused.

Creating all the tasks and resources required to execute
\figureref{fig:flowchart} and making sure that resources are optimally shared
between tasks is a complex bookkeeping exercise that can be executed very
efficiently. Once the tasks and the graph of task and resource dependencies are
constructed, we pass them on to the QuickSched
library\footnote{\url{https://gitlab.cosma.dur.ac.uk/swift/quicksched}; a
version of this library containing some important changes to make it work with
C++ is contained in the \textsc{CosTuuM} repository.} \citep{2016Gonnet} that
takes care of the parallel execution of the graph.

The various resources involved in the calculations need to be stored in memory.
In order to avoid expensive memory allocations that could lead to severe
bottlenecks during parallel execution, we preallocate all required memory during
task construction. Some of the components (e.g. the spherical Bessel functions
required for the T-matrix calculation and the T-matrices themselves) have large
sizes, which makes it impractical to allocate a unique memory block for each of
them within the memory limits of a modern system. We therefore limit the total
number of these blocks and reuse blocks that are no longer used. This is only
possible if we add additional dependencies between tasks that ensure a memory
block can only be reused if the previous tasks that depended on its old state
have finished, which further complicates the bookkeeping.

In the end, the total allowed memory usage of \textsc{CosTuuM} is treated as an
input parameter. If the allowed memory size is insufficient to fit the resources
required for a single ensemble T-matrix calculation, the calculation will abort.
The minimum required size depends on the maximum allowed spherical basis
function expansion in the T-matrix calculation and the requested accuracy, but
is typically of the order of $1-10$~GB.

\begin{figure*}
\centering{}
\includegraphics[width=0.98\textwidth]{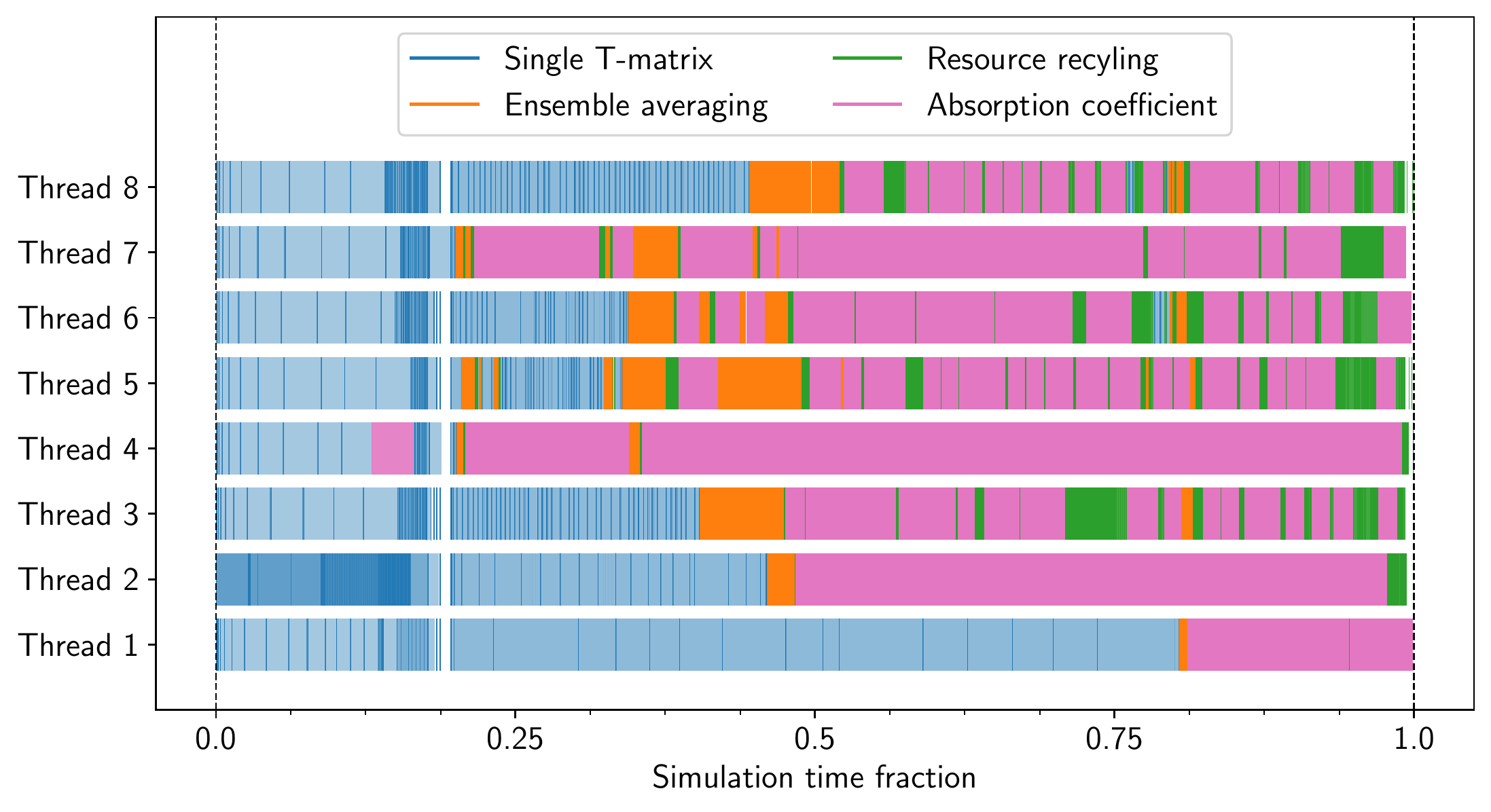}
\caption{Task plot for a small \textsc{CosTuuM} calculation using 8 computing
threads. The different colors correspond to the different steps in
\figureref{fig:flowchart}, as indicated in the legend. Some steps are not
visible because their contribution is too small; they have been omitted from
the legend. Different opacity values within the same color correspond to
different subtasks within a step. The green bars correspond to technical tasks
required to handle the reuse of resources.}
\label{fig:taskplot}
\end{figure*}

The QuickSched library internally stores hardware counters that measure the
elapsed CPU time for each task during execution. This makes it possible to
generate detailed diagnostic \emph{task plots} that show how each parallel
thread involved in the computation performs over time. \figureref{fig:taskplot}
shows an example of a task plot generated by one of our parallel runs. As can be
seen, the execution of the tasks happens stochastically and unsynchronized, with
different threads executing different tasks at the same time according to what
is available. Overall, the parallel efficiency of the algorithm is good, with
only small bottlenecks or load-imbalances (which show up as empty gaps in the
task plot).

\subsection{Performance}

\begin{figure}
\centering{}
\includegraphics[width=0.48\textwidth{}]{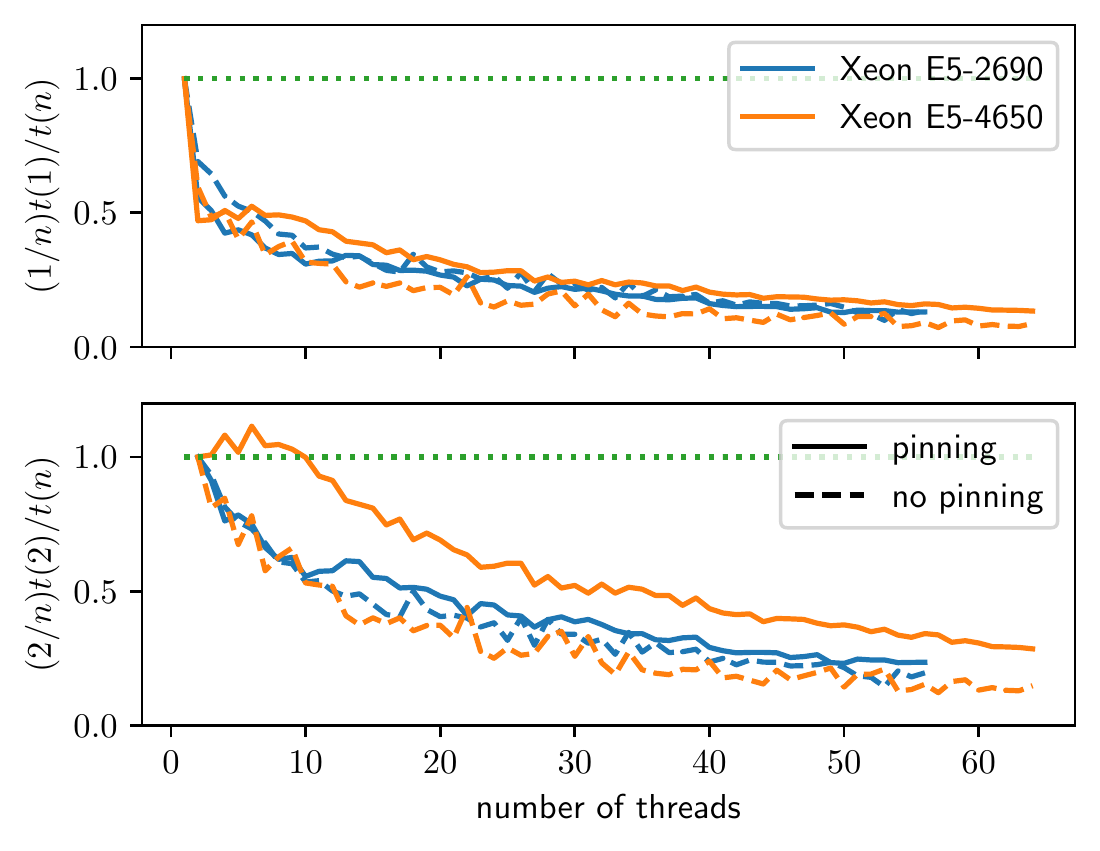}
\caption{Parallel efficiency for a strong scaling test with \textsc{CosTuuM} as
measured on two different systems using two different ways to assign parallel
threads to physical cores, as indicated in the legends. The top panel shows the
usual parallel efficiency computed using a single thread run as reference, the
bottom panel shows the same quantity but using twice the two thread run as
reference. The latter partially eliminates the effect of additional memory
costs.}
\label{fig:strong_scaling}
\end{figure}

\figureref{fig:strong_scaling} shows the parallel efficiency of \textsc{CosTuuM}
for the same run as shown in the task plot, but now for a variable number of
threads. We measured performance on two different systems: a 64 core system
using Intel Xeon E5-4650 CPUs, and a 56 core system using Intel Xeon E5-2690
CPUs. Both systems have a similar memory layout, but the latter system is
faster due to a higher CPU clock speed and better CPU optimizations. For both
systems, we ran with two different ways of assigning threads to physical cores
as set by the OpenMP OMP\_PROC\_BIND environment variable. The default value
(False) allows the kernel to run threads on the same core and move threads
during the calculation, while a value of True forces threads to run on different
cores and binds them to that core, this is called \emph{pinning}.

On both systems, there is a significant decrease in efficiency when moving from
1 to multiple threads, and this decrease is more pronounced when pinning forces
all threads to run on separate cores. This decrease is entirely due to the
memory access pattern of \textsc{CosTuuM}: a number of tasks in the task graph
make use of large blocks of continuous memory and the average execution time of
these tasks increases when increasing the number of threads from 1 to 2. If we
factor in the increase in memory access time by using the 2 thread run as a
reference to compute the parallel efficiency, then the scaling of
\textsc{CosTuuM} is significantly better. The slower system, which is less
affected by memory bandwidth issues benefits more from this approach; this
system also shows a more significant efficiency gain when threads are pinned to
cores.

We conclude that the task based algorithm does lead to low load imbalances
during parallel runs, but that the current memory usage of \textsc{CosTuuM}
hampers its overall performance. Nevertheless, a parallel speedup of up to a
factor 10 is achievable, even for small calculations. The memory issues we found
are difficult to solve and may be addressed in future versions of the library.

\bibliographystyle{aasjournal}
\bibliography{main}

\end{document}